\documentclass[prd,aps,showpacs,floats,floatfix,superscriptaddress,nofootinbib]{revtex4}

\usepackage{graphicx}
\usepackage{amsfonts} 
\usepackage{amssymb}

\newcommand{\eqref}[1]{(\ref{#1})}

\newcommand{\ket}[1]{{#1}}
\newcommand{\braket}[2]{{\langle #1,#2 \rangle}}

\begin{document}

\title{Cover art: issues in the metric-guided and metric-less placement of random and stochastic template banks}

\author{Gian Mario Manca} 

\author{Michele Vallisneri}
\affiliation{Jet Propulsion Laboratory, California 
Institute of Technology, Pasadena, CA 91109} 

\begin{abstract}
The efficient placement of signal templates in source-parameter space is a crucial requisite for exhaustive matched-filtering searches of modeled gravitational-wave sources, as well as other searches based on more general detection statistics. Unfortunately, the current placement algorithms based on regular parameter-space meshes are difficult to generalize beyond simple signal models with few parameters. Various authors have suggested that a general, flexible, yet efficient alternative can be found in randomized placement strategies such as random placement and stochastic placement, which enhances random placement by selectively rejecting templates that are too close to others. In this article we explore several theoretical and practical issues in randomized placement: the size and performance of the resulting template banks; the very general, purely geometric effects of parameter-space boundaries; the use of quasirandom (self-avoiding) number sequences; most important, the implementation of these algorithms in curved signal manifolds with and without the use of a Riemannian signal metric, which may be difficult to obtain. Specifically, we show how the metric can be replaced with a discrete triangulation-based representation of local geometry. We argue that the broad class of randomized placement algorithms offers a promising answer to many search problems, but that the specific choice of a scheme and its implementation details will still need to be fine-tuned separately for each problem.
\end{abstract}

\maketitle

\section{Introduction}
\label{sec:intro}

An international network of ground-based laser-interferometric gravitational-wave (GW) detectors \cite{ligoweb,virgoweb,geoweb}, which operate in the high-frequency band between 10 and $10^3$ Hz, has by now completed several rounds of science data collection and analysis. Space-based detectors such as LISA \cite{lisaweb} will extend our reach to frequencies as low as $10^{-5}$ Hz, and to sources as far as redshifts of $20$. The first direct detections of GW sources will inaugurate the era of GW astronomy, a novel mode of inquiry of the Universe that will provide unique information about the properties of relativistic objects such as black holes and neutron stars, and allow precise tests of general relativity's yet unproven predictions.

Because GW signals come immersed in noisy detector data, the statement that a GW source was detected is by its nature probabilistic, and can be quantified with frequentist false-alarm and false-dismissal probabilities \cite{Wainstein}, or with a Bayesian odds ratio \cite{Jaynes}. In an ideal world where the statistical properties of detector noise are known exactly and detector calibration is perfect, detection corresponds to a simple mathematical statement about the data \cite{Finn:1992p1253}, although actually evaluating that statement may require considerable numerical computation. In the real world where noise is non-Gaussian, nonstationary, and poorly characterized, and where calibration is challenging, detection corresponds to a more complicated, heuristic process, which involves expert judgment (such as the choice of data-quality vetoes and multidetector coincidence windows), and which is embodied in complex software pipelines \cite{Abbott:2009p1255}. The validity and efficiency of the process are evaluated empirically, by measuring the resulting background of spurious detection candidates, and by injecting simulated GW signals and verifying their recovery.
Indeed, the day-to-day work of GW analysts is very much dominated by the development and validation of such pipelines.

Searches for GW signals of known shape from well-understood sources, such as the archetypical gravitational two-body problem of compact-binary inspiral, merger, and ringdown, occupy a privileged spot in the GW data-analysis landscape because they are likely to yield the first robust detections, although searches for unmodeled GW bursts in coincidence with EM counterparts \cite{Abbott:2008p1256} are also a contender. Furthermore, modeled-signal searches find a straightforward mathematical realization in the simple yet powerful technique of \emph{matched filtering} \cite{Wainstein}: roughly speaking, the detector output is time-correlated with a set of theoretical waveforms (\emph{templates}), and a peak in one of the correlation time series indicates that a GW signal of that shape may be present in the data at the time of the peak. Higher peaks correspond to higher detection confidence, because stronger signals have smaller probability of being simulated by random features in the noise.

The \emph{template placement} problem arises because we are interested in searching for GW sources with a range of physical parameters (such as the component masses in a binary), and because different parameters can yield considerably different waveforms. Therefore we must choose a finite number of templates to correlate with the detector data; these templates must span the region of interest in source parameter space; they must be placed ``densely'' enough to avoid reducing detection confidence (the height of the correlation peaks) by matching a true GW signal with too different a template; yet they must be spaced as ``sparsely'' as possible to limit their total number and therefore the computational cost of the search. The notion of distance implied here by ``densely'' and ``sparsely'' is made quantitatively precise by defining a noise-weighted distance $\Delta$ [Eq.\ \eqref{eq:distance} in Sec.\ \ref{sec:matchedfiltering}] in the linear space $\mathbb{R}^{N_\mathrm{big}}$ of all possible detector outputs. The distance $\Delta$ between two GW signals is directly related to the reduction in detection confidence incurred by using one to match the other, and the noise weighting encodes the different sensitivity of the detector at different frequencies. Thus we wish to place templates uniformly from the ``point of view'' of the detector.

Now, a family of signal templates $\{h(\lambda_j)\}$ parametrized smoothly by the source parameters $\lambda_j$ realizes an $N_\mathrm{small}$-dimensional topological manifold $\mathcal{H}$ embedded in $\mathbb{R}^{N_\mathrm{big}}$. Thus the placement problem consists in selecting a finite and discrete \emph{bank} $\{h(\lambda^{(k)})\} \subset \mathcal{H}$ such that the ($\mathbb{R}^{N_\mathrm{big}}$-)distances of all possible incoming GW signals (i.e., all the points in $\mathcal{H}$) from the nearest bank template $h(\lambda^{(k)})$ satisfy certain properties. For instance, we may constrain the average and variance of this distance, or we may require that the distance may be no larger than a maximum value.
In the latter case, as emphasized by Prix \cite{Prix:2007p1257}, template placement becomes an instance of the \emph{sphere-covering} problem: given the manifold $\mathcal{H}$, we wish to cover it with the smallest possible number of spheres of fixed ($\Delta$) radius, where the sphere centers sit on $\mathcal{H}$ and correspond to the $\{h(\lambda^{(k)})\}$ in the bank. Qualitatively speaking, we want the spheres to cover the manifold completely, so that we miss no signal; at the same time we wish to reduce their overlap, so that we not waste computational resources on testing templates that are almost duplicates.

The distance $\Delta$ endows $\mathcal{H}$ with a local Riemannian structure, and therefore an $N_\mathrm{small}$-dimensional metric tensor $g_{jk}$ \cite{Balasubramanian:1996p1260,Owen:1996p1258,Owen:1999p1259}. Template placement can then be reformulated as the problem of covering $\mathcal{H}$ with spheres of fixed \emph{geodetic radius}. If $\mathcal{H}$ is curved, the geodetic distance will consistently overestimate the $\Delta$ distance, so the reformulation is only faithful when curvature is negligible over the typical sphere radius. GW analysts have generally been happy to accept this approximation and thus work in the familiar framework of differential geometry; so shall we.

If $\mathcal{H}$ has no intrinsic curvature, and if a suitable change of coordinates can be found that maps $\mathcal{H}$ to Euclidean space,\footnote{For instance, approximate but accurate mappings can be found for the 2.5PN restricted inspiral signals from binaries of spinless compact bodies \cite{Tanaka2000,Croce2001} and for the continuous GW signals from rapidly rotating neutron stars \cite{Pletsch2009}.} therefore mapping $g_{jk}$ to the identity matrix, we can rely on a wealth of theoretical results about the covering properties of \emph{periodic lattices}, and place sphere centers (i.e., bank templates) at the lattice vertices: for instance, on a square \cite{Babak:2006p1262} or hexagonal \cite{Croce:2002p1269,Cokelaer:2007p1261} lattice on the Euclidean plane, or on their generalization in higher-dimensional Euclidean space \cite{Prix:2007p1257}. If however no such change of coordinates can be found, perhaps because the metric itself is difficult to obtain or unreliable, or if $\mathcal{H}$ has significant curvature, it becomes very hard to place templates along any regular structure. In the case of two source parameters, Arnaud et al.\ \cite{Arnaud:2003p1267} and Beauville et al.\ \cite{Beauville:2003p1266} describe algorithms that create curvature-conforming tilings by propagating a locally hexagonal lattice while adjusting for the change in the equal-distance contours. Such an approach has not been attempted in higher-dimensional spaces, where it appears very cumbersome.

A possible alternative is placing templates at the vertices of suitably ``grown'' unstructured meshes, such as adaptively refined triangulations. We investigate such a strategy in a separate article \cite{vallismeshes}, and briefly in Sec.\ \ref{sec:triang} below. Here we concentrate instead on the \emph{randomized sampling} strategy originally proposed by Allen \cite{babakallen2002} and later studied by Babak \cite{babak2008}, Messenger et al.\ \cite{Messenger2009}, and Harry et al.\ \cite{Harry}.
In its simplest incarnation (pure random sampling), the idea is to draw a set of points that are (pseudo-)randomly distributed across $\mathcal{H}$. Because of its very randomness, such a covering can never be guaranteed to cover 100\% of $\mathcal{H}$; however, as shown by Messenger and colleagues \cite{Messenger2009}, it can achieve very high covering fractions with total numbers of points that (in Euclidean space) are competitive with the most efficient \emph{relaxed} lattice coverings (i.e., lattices rescaled to cover a fraction $<$ 100\% of space). To cover curved manifolds, it is sufficient to know the metric determinant, and adjust the local probability of random draws accordingly. (Recently, R\"over has suggested that the Bayesian prior distribution for the source parameters can be used to adjust the local template density in order to minimize the expected number of draws until a matching template is found \cite{roever2009}.)

In this paper we extend the results of Ref.\ \cite{Messenger2009}: we investigate the finer details of random coverings, such as the variance of the covering fraction and the effects of boundaries, and we provide further evidence that uniform coverings of curved manifolds can be obtained in general conditions, even without knowledge of the metric, but using only the $\Delta$ distance. Specifically:
\begin{itemize}
\item We confirm that random coverings are more efficient than could be naively expected, and we provide an exact formula for the variance of the covering fraction.
\item We extend Messenger et al.'s \emph{in-the-bulk} results by considering the effects of manifold boundaries in reducing the covering fraction: we find that these effects become overwhelming, and must be treated carefully, as the manifold dimension increases. For example, neglecting boundary effects, 11,000 random spheres of radius 0.3 cover the 8-dimensional unit cube with 95\% covering fraction; including boundary effects, 92,000 spheres are needed. 
\item We show that \emph{quasirandom} numbers (self-avoiding deterministic sequences often used to reduce variance in Monte Carlo integration \cite{recipes}) can be used instead of pseudorandom numbers to improve the efficiency of random coverings, although such a substitution is most effective on quasiflat manifolds.
\item We investigate \emph{stochastic} placement strategies \cite{babak2008,Harry} where the random distribution is tempered by considering the distances between selected points (e.g., rejecting draws that are too close to another point in the bank): we quantify the efficiency of these coverings, and examine their application to curved manifolds without the use of the metric.
\item Last, we propose an alternative \emph{metricless} placement strategy, based on triangulating the curved manifold to represent its local geometry as the set of finite distances between neighboring triangulation points.
\end{itemize}
Overall, we find that even flexible placement strategies such as random, stochastic, and unstructured-mesh coverings cannot be made into recipes of general applicability, but need to be tailored to the circumstances of different template families and searches. Much as the simple mathematical formalism of matched-filtering GW searches has made way to complex processes and pipelines, so must template placement ultimately evolve from a problem of mathematics to one of engineering. On the other hand, we expect that the general methods and results discussed in this paper would apply also to the larger class of GW searches that use more general detection statistics than the matched-filtering SNR \eqref{eq:snr}, but that still require the placement of (generalized) templates according to notion of distance. Examples include stack--slide \cite{brady2000}, Hough-transform \cite{krishnan2004}, $\mathcal{F}$-statistic \cite{prix2007}, and global-correlation \cite{Pletsch2009} searches for periodic GWs.

We note also that not all matched-filtering searches rely on template banks. The broad family of Monte Carlo methods, which include Markov Chain Monte Carlo, genetic algorithms, and nested sampling (see \cite{Vallisneri09} for a LISA-centric review), conceptualize searches as the randomized exploration of template--signal correlation across parameter space, rather than its evaluation at a set of predetermined points. However, as we argue in Sec.~\ref{sec:conclusions}, the systematic exploration of parameter space enabled by random and stochastic banks, as well as the scaling properties derived here and in Refs.\ \cite{Messenger2009,Harry}, may find applications in designing or guiding Monte Carlo searches.

This paper is organized as follows.
In Sec.~\ref{sec:matchedfiltering}, we recall key formulas from the theory of matched filtering, and emphasize their geometric interpretation.
In Sec.~\ref{sec:covering}, we introduce the sphere-covering problem.
In Sec.~\ref{sec:random}, we study the properties of random coverings.
In Sec.~\ref{sec:boundary}, we examine the effects of boundaries.
In Sec.~\ref{sec:alternatives}, we quantify the improved efficiency possible with quasirandom sequences and stochastic placement.
In Sec.~\ref{sec:triang}, we explore the use of unstructured meshes for metricless placement.
In Sec.~\ref{sec:conclusions}, we summarize our conclusions, and discuss future directions of investigation.
In the Appendix, we describe the numerical procedures used throughout this work.

\section{Key matched-filtering formulas}
\label{sec:matchedfiltering}

Matched-filtering searches for GW signals are based
on the systematic comparison of the measured detector output $s$
with a bank of theoretical signal templates $\{h(\lambda^{(k)})\}$.
Here we develop only the formulas that we need later, but
a few useful starting points in the vast literature on this subject are Refs.\
\cite{Wainstein,Oppe,Finn:1992p1253,Sathyaprakash:1991mt,Dhurandhar:1992mw,Cutler:1994ys,lrr-2005-3}.

\paragraph*{Inner product.}
A crucial mathematical construct is the symmetric \emph{inner
product} $\braket{g_1}{g_2}$ between two real signals (detector outputs over a fixed time period) $g_1(t)$ and
$g_2(t)$. This product is essentially the cross-correlation between
$g_1(t)$ and $g_2(t)$, weighted to emphasize the frequencies
where detector sensitivity is better. We follow Cutler and
Flanagan's conventions \cite{Cutler:1994ys} and define
\begin{equation}
\braket{g_1}{g_2} = 2 \int_{-\infty}^{+\infty} \frac{\tilde{g_1}^*(f)
\tilde{g_2}(f)}{S_n(|f|)} df =
4 \, \mathrm{Re} \int_{0}^{+\infty}
\frac{\tilde{g_1}^*(f) \tilde{g_2}(f)}{S_n(f)} df,
\label{eq:innerproduct}
\end{equation}
where the tildes denote Fourier transforms, the stars complex conjugates, and $S_n(f)$ the one-sided power spectral density of detector noise.
We can then define the
\emph{signal-to-noise ratio} (SNR) $\rho$ of detector output $\ket{s}$
after filtering by template $h(\lambda^{(k)}) \equiv h^{(k)}$ as
\begin{equation}
\rho(h^{(k)}) = \frac{\braket{s}{h^{(k)}}}{\mathrm{rms} \, \braket{n}{h^{(k)}}} =
\frac{\braket{s}{h^{(k)}}}{\sqrt{\braket{h^{(k)}}{h^{(k)}}}},
\label{eq:snr}
\end{equation}
where $n$ denotes detector noise, and $\mathrm{rms} \, \braket{n}{h^{(k)}}$ represents the rms average of that inner product over all possible noise realizations (see, e.g.,
\cite{Wainstein}).
It is convenient to think of Eq.\ \ref{eq:snr} as the SNR of $s$ after filtering by the \emph{normalized} template $\hat{h}^{(k)} \equiv h^{(k)}/\sqrt{\langle h^{(k)}, h^{(k)}\rangle}$. If $s$ consists solely of Gaussian and stationary detector noise,
then $\rho$ is a normal random variable with mean zero and variance one.
If instead a signal equal to $A \, \hat{h}^{(k)}$ is present in $s$ in addition to noise, then $\rho$ is a normal random variable with mean $A$ and again variance one.
We claim a detection of a signal with source parameters $\lambda^{(k)}$ whenever $\rho(h^{(k)})$ is greater than a chosen \emph{threshold} $\rho^*$. Such a detection scheme is Neyman--Pearson \emph{optimal} in the sense that it maximizes the probability of correct detection for a given probability
of false detection.

\paragraph*{Distance.}
Henceforth, we shall restrict our consideration to \emph{normalized} signals, and drop the caret that we just used to denote them. Given two such signals, we can use the inner product to define a \emph{distance} between them,
\begin{equation}
\Delta(g_1,g_2) = \sqrt{\braket{g_1 - g_2}{g_1 - g_2} / 2} = \sqrt{1-\braket{g_1}{g_2}}.
\label{eq:distance}
\end{equation}
This distance is a proper \emph{metric}, since it is semidefinite positive (and vanishes only for $g_1 = g_2$), symmetric, and it satisfies the triangle inequality,
\begin{equation}
\Delta(g_1,g_2) \leq \Delta(g_1,g_3) + \Delta(g_3,g_2).
\label{eq:triangle}
\end{equation} 
As mentioned in the introduction, we think of signals as points in the metric space $\mathbb{R}^{N_\mathrm{big}}$ of all possible detector outputs; $N_\mathrm{big}$ is large but finite, because the response of the detector and the frequency content of prospective GW signals are band-limited, and thus can be represented by a finite number of samples (or equivalently Fourier coefficients, or coefficients over some other basis).
Furthermore, a signal family $\{\ket{h(\lambda)}\}$ (e.g., the GW signals from  neutron-star binaries over a given range of masses) that is smoothly parametrized by the source parameters $\lambda_j$ realizes an $N_\mathrm{small}$-dimensional topological manifold embedded in $\mathbb{R}^{N_\mathrm{big}}$. The inner product endows this manifold with a Riemannian structure, and the distance between nearby points on it is then approximated by a metric tensor $g_{kl}$,
\begin{equation}
\Delta^2(h(\lambda_j),h(\lambda_j+d\lambda_j)) \approx -\frac{1}{2}
\left.\frac{d^2\braket{h(\lambda_j)}{h(\lambda'_j)}}{d \lambda'_k d \lambda'_l}\right|_{\lambda_j} \!\! d\lambda_k d\lambda_l =
g_{kl}(\lambda_j) d\lambda_k d\lambda_l.
\label{eq:metric}
\end{equation}
One criterion to choose the bank templates $\{h(\lambda^{(k)})\}$ from the continuous family $\{h(\lambda)\}$ is to require that for every signal in $\{h(\lambda)\}$, the nearest bank template be at most at a distance $\Delta_\mathrm{max}$. This ensures that the SNR is reduced \emph{at worst}\footnote{Because SNR is inversely proportional to luminosity distance, the \emph{horizon distance} to which a given source can be detected is reduced at worst by a factor MM. It follows that if sources are distributed uniformly around the detector, the rate of event detection will be reduced by a factor not worse than $\mathrm{MM}^3$.} by a factor $\mathrm{MM} = 1 - \Delta^2_\mathrm{max}$, known as the \emph{minimum match}.\footnote{However, at least in the case of binary inspiral signals, interpolation across a sparse cubic-lattice bank can be used to retrieve a fraction of SNR larger than the bank's nominal MM \cite{Croce2000,Croce2001,Mitra2005}. The gains possible with this technique are not reflected in our analysis, which is concerned with the general covering problem rather than with specific cases.}
With this criterion, template placement can be seen as an instance of the covering problem, the subject of the rest of this paper.

\paragraph*{Extrinsic parameters.}
It is sometimes possible to factor out certain source parameters (known as \emph{extrinsic}) from the template placement problem: this happens when the SNR can be maximized analytically or numerically over the appropriate range of the extrinsic parameters, so it is not necessary to explicitly place templates over a discrete set of their values. Two examples are the time of arrival and initial phase of binary inspiral signals. Extrinsic parameters are usually welcome, because they reduce the computational cost of searches; on the other hand, they complicate the geometric interpretation of template placement, because the distance between two signals is normally a function of \emph{all} source parameters, so we need a prescription to choose representative values of the extrinsic parameters.

If we denote the extrinsic parameters collectively as $\phi$, and the \emph{intrinsic} (nonextrinsic) parameters as $\theta$, the logic of matched-filtering detection would suggest using the \emph{minmax} prescription
\begin{equation}
\Delta(h(\theta_1),h(\theta_2)) \stackrel{?}{=} \sqrt{1-\min_{\phi_1} \max_{\phi_2} \braket{h(\theta_1,\phi_1)}{h(\theta_2,\phi_2)}}:
\label{eq:minmaxd}
\end{equation}
if we think of $h(\theta_1,\phi_1)$ as the GW signal to be detected, and of $h(\theta_1,\phi_1)$ as the nearest template, then $\max_{\phi_2}$ represents the ``automatic'' maximization of SNR over the extrinsic parameters, while $\min_{\phi_1}$ would conservatively ensure that the minimum match is achieved even in the least favorable case. Unfortunately, Eq.\ \eqref{eq:minmaxd} does not yield a true distance: it is not symmetric, and it does not satisfy the triangular inequality.

An alternative is to choose the extrinsic parameters as arbitrary smooth functions of the intrinsic parameters,
\begin{equation}
\Delta(h(\theta_1),h(\theta_2)) = \sqrt{1-\braket{h(\theta_1,\phi_1(\theta_1))}{h(\theta_2,\phi_2(\theta_2))}},
\label{eq:givephid}
\end{equation}
and verify that it yields an acceptable distribution of $\phi_2$-maximized SNRs over all $\phi_1$. (For instance, we may choose $\phi(\theta)$ to locally maximize the determinant of the full metric $g_{jk}$.) Throughout the rest of this paper, we shall assume that a sensible distance along the lines of \eqref{eq:givephid} can be defined whenever extrinsic parameters are present. 


\section{The covering problem: periodic lattices}
\label{sec:covering}

Prix \cite{Prix:2007p1257} discusses template placement as an instance of the sphere-covering problem. See that paper and Ref.\ \cite{Messenger2009} for a general introduction. Here we focus on essential formulas and new insights. We adopt the notation of Ref.\ \cite{Messenger2009}. 

\paragraph*{Thickness.}
The problem of covering $d$-dimensional Euclidean space $\mathbb{E}^d$ with a regular periodic lattice is well known in the mathematical literature \cite{Conway1999}:
it consists in finding the lattice with the smallest number of points per unit volume such that, if a $d$-dimensional sphere of fixed radius $r$ is placed with its center at each lattice point, the spheres cover $\mathbb{E}^d$ completely.  A typical quantity used to evaluate the efficiency of such a covering is the \emph{thickness} $\Theta$, defined 
(for a subvolume $V_m$ of $\mathbb{E}^d$) as
\begin{equation}
\Theta=\frac{N \, V_d \, r^d }{V_m},
\label{eq:thickness}
\end{equation}
where $N$ is the number of lattice points (and therefore spheres), $V_d$ the volume of the $d$-dimensional unit-radius sphere, and $r$ the required sphere radius. The unit-sphere volume $V_d$ can be expressed in terms of the gamma function as $V_d=\pi^{d/2} / \Gamma(d/2+1)$; its values for $d$ up to 20 are given in Table \ref{Tab:dsphere}.
\begin{table}
\begin{tabular}{rr|rr|rr|rr}
\hline \hline
$d$ & $V_d$ & $d$ & $V_d$ & $d$ & $V_d$ & $d$ & $V_d$ \\
\hline
1 & 2.00 &  6 & 5.17 & 11 & 1.88 & 16 & 0.24 \\
2 & 3.14 &  7 & 4.72 & 12 & 1.34 & 17 & 0.14 \\
3 & 4.19 &  8 & 4.06 & 13 & 0.91 & 18 & 0.08 \\
4 & 4.93 &  9 & 3.30 & 14 & 0.60 & 19 & 0.05 \\
5 & 5.26 & 10 & 2.55 & 15 & 0.38 & 20 & 0.03 \\
\hline \hline
\end{tabular}
\caption{Volume of the $d$-dimensional unit-radius sphere. Interestingly, $V_d$ is maximum for $d=5$.}
\label{Tab:dsphere}
\end{table}
\begin{table}
\begin{tabular}{r|ll|rrrr|rrrr|rrrr}
\hline \hline
$d$ & \multicolumn{2}{c|}{Complete} & \multicolumn{4}{c|}{Partial, $(\mathrm{A}d)^*$ lattice} & \multicolumn{4}{c|}{Partial, $\mathbb{Z}^d$ lattice} & \multicolumn{4}{c}{Random} \\
 & $\Theta_{100\%}$ & best known & $\Theta^{(\mathrm{A}d)^*}_{90\%}$ & $\Theta^{(\mathrm{A}d)^*}_{95\%}$& $\Theta^{(\mathrm{A}d)^*}_{99\%}$ & $\Theta^{(\mathrm{A}d)^*}_{100\%}$ &$\Theta^{\mathbb{Z}^d}_{90\%}$ & $\Theta^{\mathbb{Z}^d}_{95\%}$ &  $\Theta^{\mathbb{Z}^d}_{99\%}$ & $\Theta^{\mathbb{Z}^d}_{100\%}$ & $\Theta_{90\%}$ & $\Theta_{95\%}$& $\Theta_{99\%}$ & $\Theta_{100\%}$ \\
\hline						     	      
1   &	1.00  & $\mathrm{Z} \equiv (\mathrm{A}1)^*$  & 0.90 & 0.95 & 0.99 &   1.00	    &  0.90 & 0.95   &   0.99 & 1.00     \\
2   &	1.21  &	$\mathrm{A}2 \equiv (\mathrm{A}2)^*$    & 0.90 & 0.97 & 1.09 &   1.21    &  0.98 & 1.13   &   1.36 & 1.57    \\
3   &	1.46  &	$(\mathrm{A}3)^*$	      & 0.97 & 1.07 & 1.21 &   1.46    &  1.16 & 1.38   &   1.82 & 2.72    \\
4   &	1.77  &	$(\mathrm{A}4)^*$	      & 1.05 & 1.18 & 1.38 &   1.77    &  1.40 & 1.74   &   2.44 & 4.93    \\
5   &	2.12  &	$(\mathrm{A}5)^*$	      & 1.16 & 1.32 & 1.57 &   2.12    &  1.72 & 2.20   &   3.30 & 9.20    \\
6   &	2.46  &	L6c1	      & 1.30 & 1.49 & 1.80 &   2.55    &  2.11 & 2.79   &   4.43 & 17.4    \\
7   &	2.90  &	L7c	          & 1.45 & 1.69 & 2.07 &   3.06    &  2.60 & 3.56   &   5.96 & 33.5    \\
8   &	3.14  &	L8c	          & 1.63 & 1.92 & 2.38 &   3.67    &  3.23 & 4.53   &   7.99 & 64.9    \\
9   &	4.27  &	L9c	          & \it{1.85}  & \it{2.18}  & \it{ 2.73}  &   4.39    &  4.00 & 5.76   &  10.6  & 127 & \multicolumn{4}{c}{(any $d$)} \\
10  &	5.15  &	L10c	      & \it{2.09}  & \it{2.50}  & \it{ 3.32}  &   5.25    &  4.97 & 7.33   &  14.2  & 249 & 2.3  & 3.0  & 4.6  &  $\infty$    \\
11  &	5.51  &	L11c	      & \it{2.45}  & \it{2.83}  & \it{ 3.77}  &   6.28    &  6.16 & 9.31   &  18.9  & 491     \\
12  &	7.47  &	L12c	      & \it{2.67}  & \it{3.34}  & \it{ 4.27}  &   7.51    &  7.67 & 11.9   &  25.0  & 973     \\
13  &	7.76  &	L13c	      & \it{3.10}  & \it{3.82}  & \it{ 5.01}  &   8.98    &  9.50 & 15.0   &  33.0  & 1930    \\
14  &	8.83  &	L14c	      & \it{3.54}  & \it{4.37}  & \it{ 5.81}  &  10.73    & 11.8  & 19.1   &  43.5  & 3860    \\
15  &   11.00 &	L15c	    & \it{4.20}  & \it{4.96}  & \it{ 6.87}  &  12.82    & 14.7  & 24.2   &  57.0  & 7700   \\
16  &	15.31 &	$(\mathrm{A}16)^*$	      & \it{4.71}  & \it{5.88}  & \it{ 7.76}  &  15.31    &...    & ...    & ...    & ...    \\
17  &	12.36 &	A17	      & \it{5.36}  & \it{6.63}  & \it{ 9.16}  &  18.29    &...    & ...    & ...    & ...    \\
18  &	21.84 &	$(\mathrm{A}18)^*$	      & \it{6.16}  & \it{7.80}  & \it{10.76}  &  21.84    &...    & ...    & ...    & ...    \\
19  &	21.23 &	A19	      & \it{6.99}  & \it{8.95}  & \it{12.49}  &  26.08    &...    & ...    & ...    & ...    \\
20  &	20.37 &	A20        & ...  & ...  & ...  &  31.14    &...    & ...    & ...    & ...    \\
\hline \hline
\end{tabular}
\caption{A comparison of thickness as function of $d$ for the most efficient known complete coverings (column 2), for partial and complete $(\mathrm{A}d)^*$ coverings (columns 4--7), for partial and complete $\mathbb{Z}^d$ coverings (columns 8--11), and for random coverings (columns 12--15). In this table, thicknesses are computed in the limit $V_m \rightarrow \infty$, neglecting any boundary effects. See also~\cite{coveringweb2,Prix:2007ks}. All values were computed using the technique discussed in the Appendix, except for those in italic, which are taken from Ref.\ \cite{Messenger2009}, and reported here for completeness.\label{Tab:thickness}}
\end{table}

The thickness is a pure number: it is the ratio between the total volume of the spheres used in the covering and the total covered volume. The best possible case is a covering with  
$\Theta=1$, where the spheres do not intersect, and each point of $\mathbb{E}^d$ is inside one and only one sphere; of course
it is impossible to reach this limit for $d > 1$. Mathematicians have worked out theoretical limits for the thinnest possible coverings, such as the CFR bound \cite{Conway1999}
\begin{equation}
\Theta>\frac{d}{e \sqrt{e}} \simeq \frac{d}{4.4817}.
\end{equation} 
The most efficient known lattice coverings for $d \leq 20$ are listed with their $\Theta$ in columns 2 and 3 of Table \ref{Tab:thickness}. Note that the authors of Ref.\ \cite{Messenger2009} usually report values for the \emph{normalized thickness} $\theta=\Theta / V_d$ rather than for $\Theta$, as we do.

\paragraph*{Partial covering.}
A useful generalization of the problem is to allow a fraction of space to remain uncovered, and define an \emph{effective thickness} that depends not only on the choice of lattice, but also on the percentage of the volume that it covers.
In the language of template banks, this means that the maximum-distance criterion will not be achieved for a fraction (usually small) of the possible incoming GW signals. As noticed in Ref.\ \cite{Messenger2009}, the effective thickness of partial regular coverings (even for rather large covering fractions) improves dramatically compared to complete coverings, especially as $d$ increases. This means that in a complete covering most of the lattice density is needed to reach the ``hardest'' few percent of the volume; by contrast, even a simple $\mathbb{Z}^d$ (simple hypercubic) lattice can efficiently cover 95\% or 99\% of space. See columns 4--11 of Table \ref{Tab:thickness} for the partial-covering thicknesses of $\mathbb{Z}^d$ and $(\mathrm{A}d)^*$ lattices.
(The latter are the $d$-dimensional generalization of body-centered cubic lattices, whereas the $\mathrm{A}d$ lattices are the generalization of face-centered cubic lattices. Here the star denotes a \emph{reciprocal} lattice. A2 is the hexagonal lattice, and is its own reciprocal. See Ref.\ \cite{Conway1999} for details.)
\begin{figure}[t]
\includegraphics[width=\textwidth]{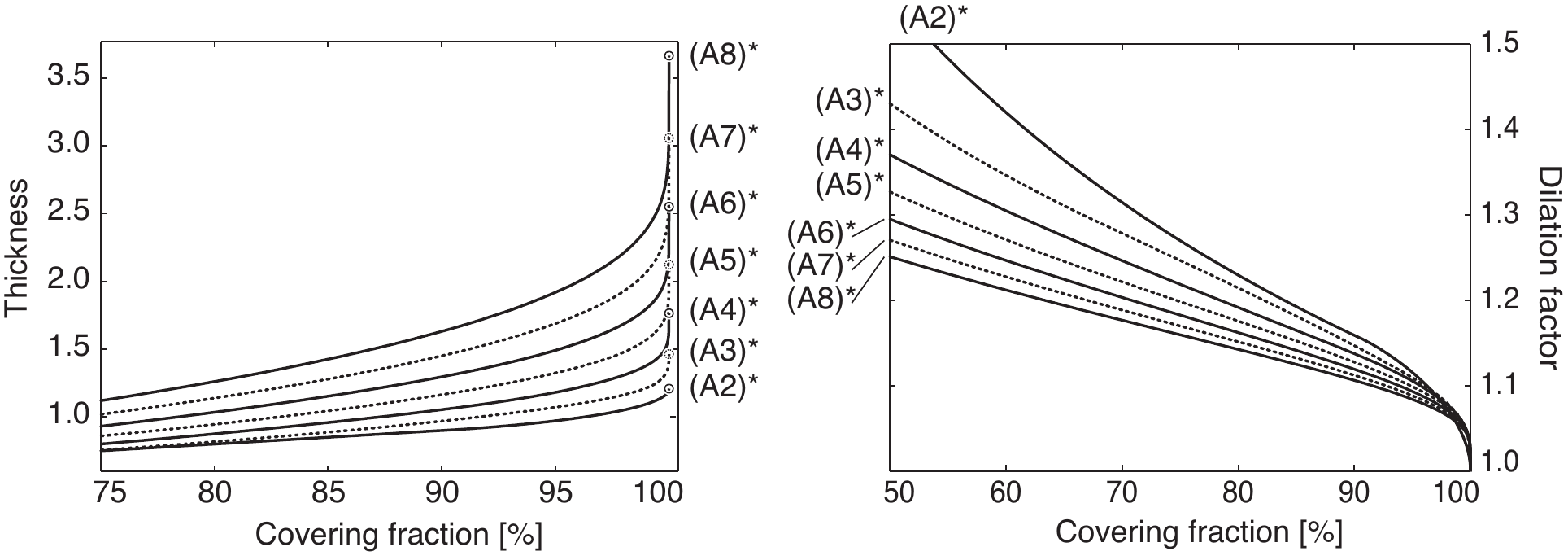}
\caption{Efficiency of partial coverings. Left: thickness of $(\mathrm{A}d)^*$ coverings as a function of the covering fraction, for $d = 2\mbox{--}8$. Right: isotropic dilation factor between a complete covering and partial covering of the same thickness, for the covering fraction given on the horizontal axis.\label{Fig:partial_An}} 
\end{figure}

In this context, a better index of covering efficiency is not the complete-covering thickness $\Theta_{100\%}$, but the variation of $\Theta$ as a function of the covering fraction. The left panel of Fig.\ \ref{Fig:partial_An} shows this function for $(\mathrm{A}d)^*$ lattices with $d = 2\mbox{--}8$. The right panel shows the isotropic dilation factor that must be applied to a complete covering to achieve a lower covering fraction with the same thickness. Thus we could say (for instance) that an $(\mathrm{A}8)^*$ covering is $\sim 2.3$ times more efficient 
at an 80\% covering fraction than at 100\%, where the factor 2.3 is the ratio of the thicknesses, which is also equal to the eighth power of the dilation factor.

\begin{figure}
\includegraphics[width=0.75\textwidth]{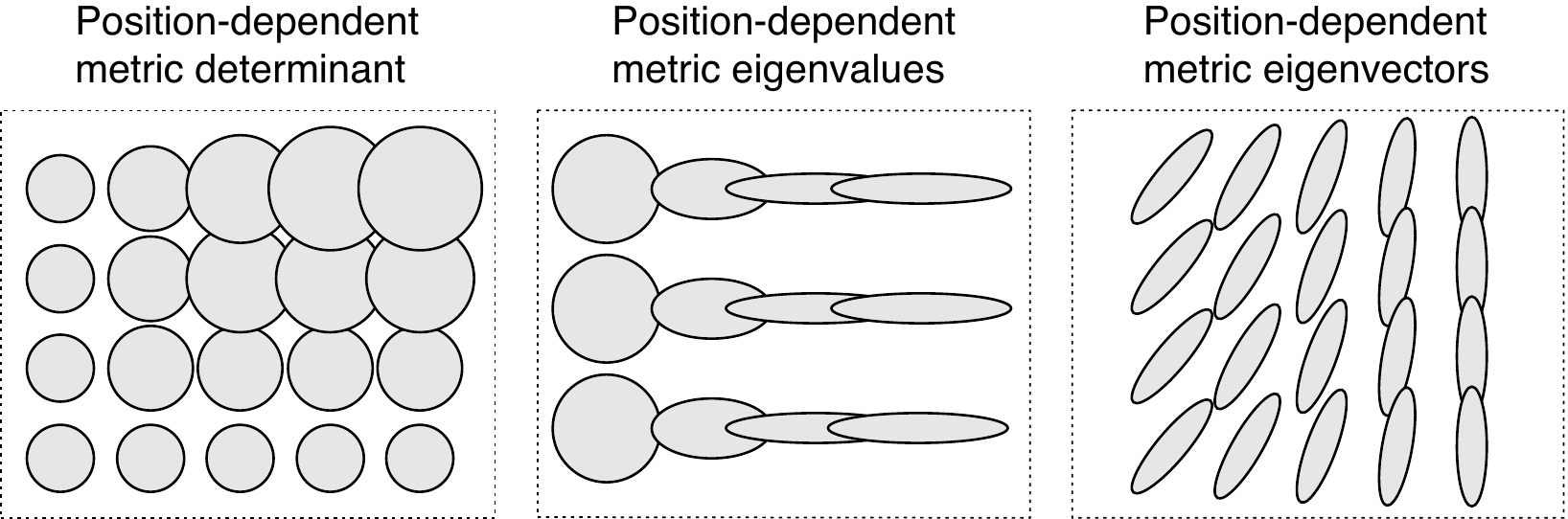}
\caption{Three ways in which manifold curvature disturbs a lattice covering laid out periodically along manifold coordinates. Left: variation in the metric's determinant causes spheres with the same geodesic-distance radius to have different coordinate radii. Center: variation in the ratio of the metric's eigenvalues maps geodesic-distance spheres into coordinate ellipses. Right: variation in the metric's eigenvectors changes the orientation of coordinate ellipses.
Because of all three effects, the covering fraction comes to depend on the position in the manifold.
\label{Fig:Illustration}}
\end{figure}

\paragraph*{Curved manifolds.}
So far we have considered the covering problem for Euclidean space, and equivalently for flat manifolds $\mathcal{H}$ that can be mapped to $\mathbb{E}^d$ by a coordinate change. What about curved manifolds?
For these, the metric $g_{kl}$ is a function of position on the manifold, and the efficiency of lattice coverings that are periodic with respect to the coordinates drops considerably. This is because of three different effects, illustrated in Fig.~\ref{Fig:Illustration}: the variation of the determinant of the metric, the variation in the ratio of its eigenvalues, and the variation in the orientation of its eigenvectors. This distinction is somewhat academic at this point, but it will acquire a deeper meaning in later sections. 

Let us first consider the effects of metric-determinant variation. 
We may refer to the right panel of Fig.~\ref{Fig:partial_An} to gain a quantitative understanding in the case of $(\mathrm{A}d)^*$ lattices. Suppose that one such lattice was arranged to provide 100\% covering around the point $\hat{\lambda}_i$ where $|g_{kl}|$ is minimum: around a different point $\lambda_i$, the lattice will be in effect dilated isotropically by a factor $(|g_{kl}(\lambda_i)|/|g_{kl}(\hat{\lambda}_i)|)^{1/2d}$; the corresponding reduction in the covering fraction can be read off from Fig.~\ref{Fig:partial_An}. Changes of a few tens \% in $|g_{kl}|$ are already very damaging.

\begin{figure}
\includegraphics[width=\textwidth]{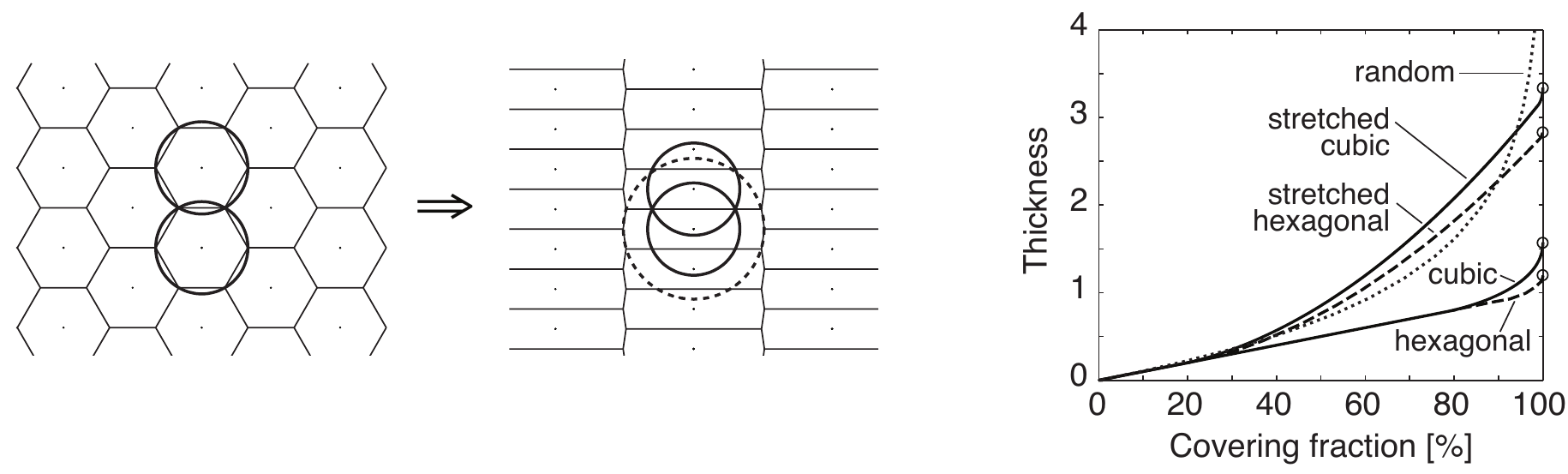}
\caption{Left: effect of dilating space along the $x$ direction and contracting along $y$, for a $d=2$ hexagonal covering. The dilation--contraction factor is 2, so the determinant of the transformation is 1. The spheres centered on the lattice points are now too far apart along the $x$ direction, too close along $y$. The new lattice realizes a covering with the larger radius (i.e., maximum template distance) of the dashed circle. Notice that the Voronoi cells (i.e., the sets of points closer to a given lattice point than to any other lattice point) of the stretched lattice are \emph{not} the stretched Voronoi cells of the original lattice. Right: reduction in covering fraction (for a fixed thickness) or increase in thickness (for a fixed covering fraction) for the dilation--contraction of the left panel, shown here also for a $d=2$ cubic covering. As we shall see in Sec.\ \ref{sec:random}, random coverings are not affected.} 
\label{Fig:dilatation_voronoi}
\label{Fig:dilatation}
\end{figure}

Let us now consider the effects of variation in the ratio of metric eigenvalues. This can be interpreted as 
a local dilation and contraction of space along orthogonal axes, as displayed in the left panel of Fig.~\ref{Fig:dilatation} for an $(\mathrm{A}2)^*$ (hexagonal) covering.
The covering becomes insufficient along one direction, and inefficient along the other. The right panel of Fig.~\ref{Fig:dilatation} shows how the dilation--contraction affects the covering-fraction--thickness curve.
The worsening covering performance can always be seen from two opposite viewpoints: either as a loss of efficiency (i.e., an increase in thickness) for the same covering fraction, or a decrease in covering fraction for the same thickness.

In fact, the thickness of a periodic lattice can be made arbitrarily large even while keeping $|g_{kl}|$ unchanged. The reason is that the points of periodic lattices lie on hyperplanes, since they can all be obtained as linear combinations with integer coefficients of $d$ independent vectors \cite{Conway1999}. Consider now the transformation given by a simultaneous contraction inside the hyperplanes defined by $d - 1$ of those vectors, and a dilation in the orthogonal direction. (Arrange the dilation--contraction factors to cancel out in $|g_{kl}|$.) Much like what happened in Fig.~\ref{Fig:partial_An}, the transformation increases the sphere radius needed to cover the space between hyperplanes, but also the superposition of the spheres along directions contained in the hyperplanes.

\section{The covering problem: random coverings}
\label{sec:random}

In Euclidean space, a random covering is obtained by choosing $N$ points randomly and uniformly distributed across the region to be covered. Such an arrangement can never be guaranteed to cover 100\% of points in all of its realizations; random coverings however generate very efficient \emph{partial} coverings of $\mathbb{E}^d$---indeed, for sufficiently high $d$, the \emph{most} efficient partial coverings \cite{Messenger2009}.

\paragraph*{Effective thickness.}
We can use probabilistic reasoning to characterize $\Theta$ for a random covering.
Assume that the probability of choosing any point in $\mathbb{E}^d$ is the same; given a (generic) reference point $P$, the probability that a sphere of radius $r$ with randomly chosen center will contain $P$ is
\begin{equation}
p_\mathrm{in} = V_d r^d / V_m < 1,
\end{equation}
the ratio of the volume of a sphere to the manifold volume to be covered. The probability that $N$ randomly centered spheres will \emph{not} contain $P$ is then 
\begin{equation}
p_N = (1 - V_d r^d / V_m)^N < 1,
\end{equation}
so the probability that \emph{at least one} of the $N$ spheres covers $P$ is $p_C = 1 - p_N$. Taking the logarithm, if the spheres are small enough compared to $V_m$ (more precisely, for $V_m$ and $N \rightarrow \infty$ with constant $V_m/N$),
\begin{equation}
\log(1 - p_C) = N \log(1 - V_d r^d / V_m) \simeq - N V_d r^d / V_m= -\Theta
\label{eq:pc}
\end{equation}
[according to definition \eqref{eq:thickness}]. Thus the effective thickness is a function of the probability that a generic point is covered,
\begin{equation}
\Theta^{\mathrm{random}}_{p_C} \simeq \log \left( \frac{1}{1-p_C} \right).
\label{eq:thetafromp}
\end{equation}
As shown in the rightmost ``Random'' section of Table \ref{Tab:thickness}, $\Theta^\mathrm{random}_{100\%}=\infty$, $\Theta^\mathrm{random}_{99\%}=4.6$, $\Theta^\mathrm{random}_{95\%}=3.0$, and $\Theta^\mathrm{random}_{90\%}=2.3$, independently of the dimension $d$.
These numbers assume the limit $r \rightarrow 0$, not just because of the approximation in Eq.\ \eqref{eq:pc}, but also to cancel out boundary effects. As we shall see below, these become critical for large $d$, or when one or more dimensions have extents comparable to $r$. Strictly speaking, $\Theta^\mathrm{random}_{95\%}$ is a function of $p_C$ (the probability that a generic point is covered), and not of the covering fraction $\mathcal{C}$, which is a random variable that depends on the particular realization of the covering. However, we shall see shortly that the average of $\mathcal{C}$ over realizations is just $p_C$, while its variance vanishes in the limit of $N \rightarrow \infty$ (and therefore of $r \rightarrow 0$).

Equation \eqref{eq:thetafromp} yields an estimate of the number of random points needed to achieve a given $p_C$. For instance, for $p_C = 95\%$,
\begin{equation}
N=\Theta^\mathrm{random}_{95\%} \left( \frac{V_m}{V_d r^d} \right) \simeq \log \left( \frac{1}{1-0.95} \right) \left( \frac{V_m}{V_d r^d} \right) \simeq 3.0 \left( \frac{V_m}{V_d r^d} \right).
\label{eq:exampleN}
\end{equation}
Let us also establish (for future use) an expression for the probability that at least one of $N$ randomly chosen points falls inside a region of volume $V_R$. Following the same line of reasoning as above, for $N \gg 1$ we find
\begin{equation}
p_\mathrm{inside} = 1-(1-V_R / V^m)^N = 1-\left( 1-\frac{\Theta}{N} \frac{V_R}{V_d r^d} \right)^N
\simeq 1-\exp \left(- \Theta \frac{V_R}{V_d r^d} \right).
\label{eq:pinside}
\end{equation}

\paragraph*{Average and variance of the covering fraction.}
Following Messenger and colleagues \cite{Messenger2009}, let us now confirm that the average covering fraction is $p_C$; extending their derivation, we shall also obtain an exact expression for the variance of the covering fraction.

We define the characteristic function $f(P)$ to be one for points that are covered in a realization of the covering, and zero for points that are left uncovered. The covering fraction is then $\mathcal{C} = (1/V_m) \int f(P) dV_m$ (where $P$ is parametrized by $\lambda_i$, and $dV_m=\sqrt{|g_{kl}(\lambda_i)|} d\lambda_i$) and its expectation value is
\begin{equation}
E[\mathcal{C}] = \frac{1}{V_m} \! \int E[f(P)] dV_m = \frac{1}{V_m} \int p_C dV_m = p_C.
\end{equation}
The computation of the variance is more involved:
\begin{figure}
\includegraphics{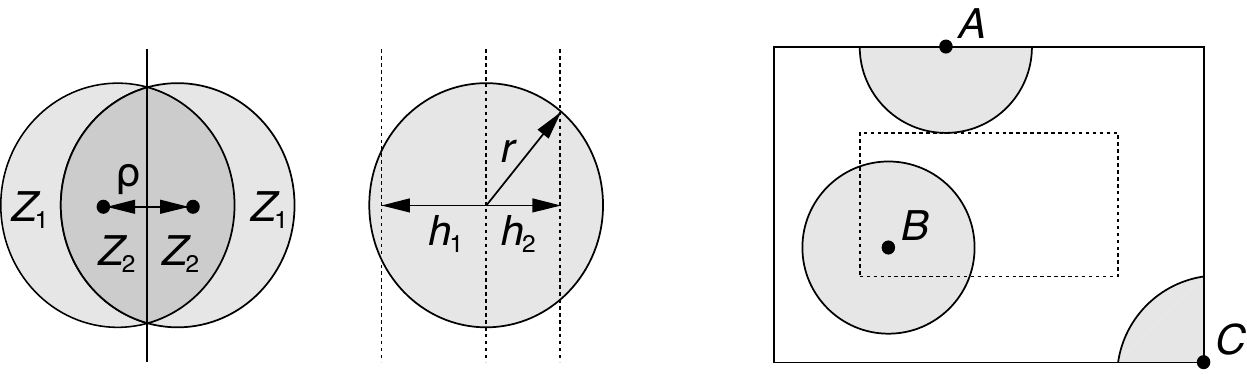}
\caption{Left: typical geometry that enters the computation of the variance of the covering fraction and of boundary effects. The basic ingredient is the volume of the slice of a sphere cut by parallel hyperplanes. Right: boundary effects. The probability that a generic point is covered scales with the ``available'' manifold volume within a distance smaller than the covering radius.\label{Fig:correlations_borders}}
\end{figure}
\begin{equation}
E[\mathcal{C}^2] - E[\mathcal{C}]^2 = 
\frac{1}{V_m^2} \! \int \!\! \int E[ f(P) f(P') ] dV_m dV_m' - p^2_C =
\frac{1}{V_m} \! \int \left( E[f(P_0) f(P)] - p^2_C \right) dV_m.
\end{equation}
If points $P_0$ and $P$ are separated by a distance larger than $2r$ (twice the covering radius), no sphere can cover both, so the probability that $P_0$ is covered and the probability that $P$ is covered are uncorrelated, $E[f(P_0) f(P)] = p^2_C$. Switching to a set of Euclidean, spherical coordinates centered in $P_0$, we then have
\begin{equation}
E[\mathcal{C}^2] - E[\mathcal{C}]^2 = 
\frac{S_d}{V_m} \! \int_{\rho<2r} \rho^{d-1} \left( E[f(0)f(\rho)] - p^2_C \right) d\rho,
\end{equation}
where $S_d$ is the surface of the $d$-dimensional unit sphere.
We define $p_\mathrm{both}(\rho) \equiv E[f(0)f(\rho)]$ to be the probability that two points separated by distance $\rho$ are both covered. Looking at the left panel of Fig.\ \ref{Fig:correlations_borders}, we conclude that
$p_\mathrm{both}=1-(1-p_{2Z_2})(1-p_{Z_1}^2)$, where $p_{2Z_2}$ is the probability of having at least one covering point in \emph{either} of the regions labeled $Z_2$, while $p_{Z_1}^2$ is the probability of having at least one point in \emph{both} the regions labeled $Z_1$. 
Now, the volumes of $Z_1$ and $Z_2$ can be expressed in terms of the volume covered by the portion of a $d$-dimensional sphere (of radius $r$, the covering radius) that is cut by parallel hyperplanes at (signed) distances $h_1$ and $h_2$ from the center (see again the left panel of Fig.\ \ref{Fig:correlations_borders}),
\begin{equation}
V_d^{\|}(h_1,h_2,r) = V_{d-1} r^d \! \int_{\arcsin(h_1/r)}^{\arcsin(h_2/r)}\cos^d(\theta) d \theta;
\label{eq:vslice}
\end{equation}
we then have 
\begin{equation}
V_{Z_2}(\rho) = V_d^{\|}(\rho/2,r,r), \quad V_{Z_1}(\rho) = r^d V_d - 2\, V_{Z_2}(\rho).
\end{equation}
Using Eq.\ \eqref{eq:pinside} to express $p_{2Z_2}$ and $p_{Z_1}$, after some manipulation we find
\begin{equation}
p_\mathrm{both}(\rho) = 1 - 2 \exp(-\Theta) + \exp \left(-2 \Theta + 2 \Theta \frac{V_{d-1}}{V_d} \! \int_{\arcsin(\rho/2r)}^{\pi/2} \cos^d(\theta) d\theta \right),
\end{equation}
 and finally
\begin{equation}
E[\mathcal{C}^2] - E[\mathcal{C}]^2 =
\frac{d \, \Theta}{N} \int_{0}^{2} y^{d-1} \left[ p_\mathrm{both}(r y) - p_C^2 \right] dy,
\label{eq:variancefinal}
\end{equation}
where we have used the fact that $\Theta/N = V_d r^d /V^m$ and that $S_d=d V_d$. Note that Eq.\ \eqref{eq:variancefinal} does not depend on the covering radius $r$. Table
\ref{Table:random_variance} gives a few numerical values for this expression, which can
be used instead of the ``guesstimate'' of Ref.\ \cite{Messenger2009},
$E[\mathcal{C}^2] - E[\mathcal{C}]^2 \sim p_C(1-p_C) / (2 \, d \, N)$. The two expressions share the initial decrease of variance with increasing $d$, and the dominant $1/N$ scaling.
\begin{table}
\begin{tabular}{r|lllll||r|lllll}
\hline \hline
\multicolumn{1}{c|}{$d$} & \multicolumn{1}{c}{50\%}   &   \multicolumn{1}{c}{80\%}    & \multicolumn{1}{c}{90\%}   & \multicolumn{1}{c}{95\%}   & \multicolumn{1}{c||}{99\%} &
\multicolumn{1}{c|}{$d$} & \multicolumn{1}{c}{50\%}   &   \multicolumn{1}{c}{80\%}    & \multicolumn{1}{c}{90\%}   & \multicolumn{1}{c}{95\%}   & \multicolumn{1}{c}{99\%}   \\
\hline
1 & 0.1534 &   0.1913  & 0.1340 & 0.0800 & 0.0189  &  11 & 0.1212 &   0.1059  & 0.0548 & 0.0234 & 0.0023  \\
2 & 0.1421 &   0.1578  & 0.1003 & 0.0540 & 0.0099  &  12 & 0.1210 &   0.1053  & 0.0543 & 0.0232 & 0.0022  \\
3 & 0.1351 &   0.1389  & 0.0826 & 0.0413 & 0.0062  &  13 & 0.1207 &   0.1049  & 0.0540 & 0.0230 & 0.0022  \\
4 & 0.1306 &   0.1274  & 0.0723 & 0.0343 & 0.0045  &  14 & 0.1206 &   0.1046  & 0.0537 & 0.0228 & 0.0022  \\
5 & 0.1276 &   0.1201  & 0.0661 & 0.0303 & 0.0036  &  15 & 0.1205 &   0.1043  & 0.0535 & 0.0227 & 0.0022  \\
6 & 0.1255 &   0.1152  & 0.0620 & 0.0278 & 0.0030  &  16 & 0.1204 &   0.1041  & 0.0534 & 0.0227 & 0.0022  \\
7 & 0.1240 &   0.1119  & 0.0594 & 0.0261 & 0.0027  &  17 & 0.1203 &   0.1040  & 0.0533 & 0.0226 & 0.0021  \\
8 & 0.1229 &   0.1096  & 0.0576 & 0.0250 & 0.0025  &  18 & 0.1203 &   0.1039  & 0.0532 & 0.0226 & 0.0021  \\
9 & 0.1222 &   0.1079  & 0.0563 & 0.0243 & 0.0024  &  19 & 0.1202 &   0.1038  & 0.0532 & 0.0225 & 0.0021  \\
10 & 0.1216 &   0.1068  & 0.0554 & 0.0238 & 0.0023 &  20 & 0.1202 &   0.1038  & 0.0531 & 0.0225 & 0.0021  \\
\hline\hline
\end{tabular}
\caption{Variance of the covering fraction [Eq.\ \eqref{eq:variancefinal}], before scaling by $1/N$. Clearly the covering fraction converges very quickly to $p_C$ for even moderate $N$. The dependence on $d$ is mild.}
\label{Table:random_variance}
\end{table}

\paragraph*{Maximum distance.}
Another obvious indicator of the performance of random coverings is the maximum distance $r_w$ of a point in the covered manifold region from the nearest covering point. This quantity coincides with the radius of the largest empty sphere that can be fit among the covering points, and again it is a random variable that depends on the particular covering realization. The authors of Ref.\ \cite{Messenger2009} make another guesstimate for the probability distribution of $r_w$:
in a given realization of a random bank in $d$ dimension, with $N$ templates and a covering radius $r$ the probability that maximum distance is smaller
than $\bar{r}$ would be
\begin{equation}
p(r_w<\bar{r}) \simeq \left[ 1 - \exp \bigl(\Theta (\bar{r}/r)^d \bigr) \right]^{2 d N}.
\end{equation}
The guesstimate fits reasonably the results of low-dimensional numerical experiments with random coverings at 90\% covering percentage.
Anyway, since this probability distribution is based on the same approximation used for the guesstimate of the variance of the covering fraction, it must be used \textit{cum grano salis}:
it may not provide, as claimed, 10\%-accurate predictions at higher $d$, or with different covering fractions.
Still, the Monte Carlo simulations presented in Ref.\ \cite{Messenger2009} do indicate that the mean and variance of $r_w$ decrease with increasing $d$, but it is not clear whether they tend asymptotically to zero (a complete covering in every realization).

\paragraph*{Curved manifolds.}
If we leave Euclidean space, the proper definition of ``random'' is that the probability for each point in the covering to lie within a manifold subregion should be proportional to the volume of that subregion. The derivation that led to Eqs.\ \eqref{eq:thetafromp} and \eqref{eq:exampleN} can then be reproduced verbatim, with the caveat that $r$ must be small enough compared to the curvature scale for $V_d r^d$ to be a good approximation to the volume of a sphere of radius $r$ (it is hard to do without this assumption, which we shall maintain to be valid in the rest of this paper).

In coordinate space, random manifold points are distributed with local density proportional to $\sqrt{|g_{kl}|}$. To achieve this in practice, we may for instance use a \emph{rejection method} whereby points $\lambda_i$ are chosen uniformly in coordinate space, and then accepted with probability $\sqrt{|g_{kl}(\lambda_i)|}/(\max_{\lambda_i} \sqrt{|g_{kl}(\lambda_i)|})$.
The computational cost of this procedure will be dominated by the estimation of the total manifold volume through the integration of $\sqrt{|g_{kl}|}$, and by its local evaluation. Thanks to the modulation of (coordinatewise) density, and to the fact that random coverings are distributed along no preferred hyperplanes, and with no preferred lengthscales, random coverings on curved manifolds avoid all three problems illustrated in Fig.\ \ref{Fig:Illustration}, and \emph{achieve the same thickness} (for the same covering fraction) as in Euclidean space. 

\section{Boundary effects for random coverings}
\label{sec:boundary}

The analysis of Messenger and colleagues \cite{Messenger2009}, as well as our discussion so far, have dealt with the thickness of coverings \emph{in the bulk}, and neglected boundary effects. Unfortunately, these can be very important, and can be difficult to correct in practice. For random coverings, it is relatively easy to understand their origin and estimate their magnitude. As illustrated for points $A$ and $C$ in the right panel of Fig.~\ref{Fig:correlations_borders}, at distances $< r$ from the boundary, the probability that a generic point is covered is smaller than farther away in the bulk; this is because the probability scales with the manifold volume contained within a radius $r$ from the point. 

\begin{figure}
\includegraphics[width=\textwidth]{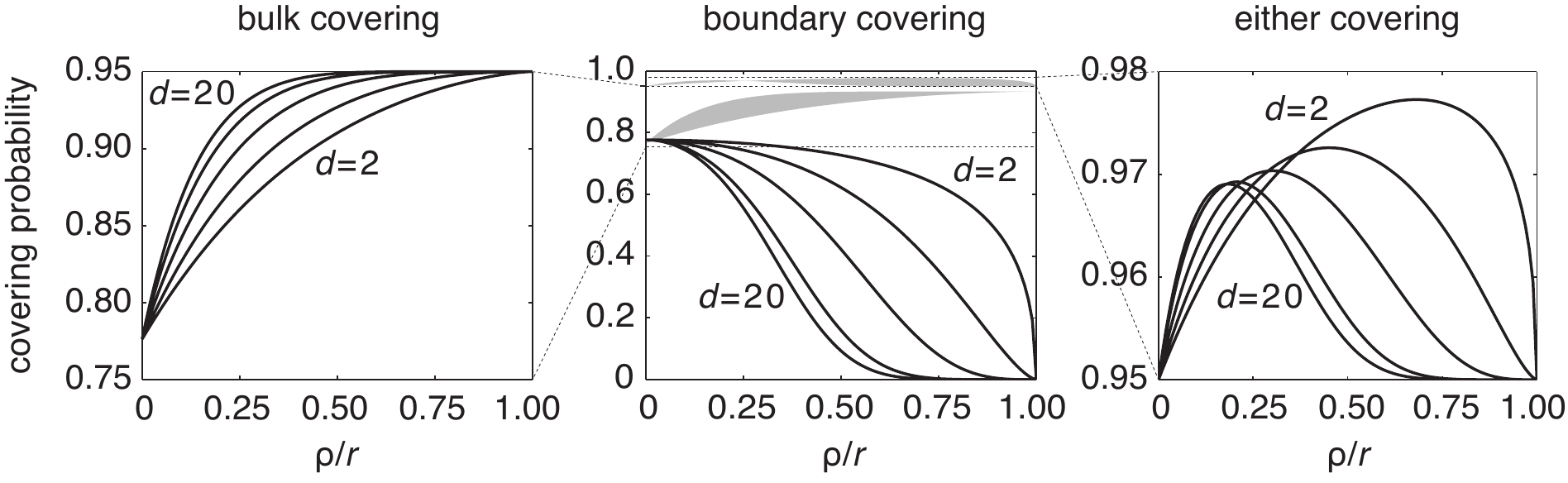}
\caption{Probability of covering a point at a distance $\rho$ from a $(d-1)$-dimensional boundary with a sphere from: left---a random bulk covering with $p_C = 95\%$; center---a random boundary covering with $p_S = 77\%$; right---either covering. As in the main text, here $r$ is the covering radius.
\label{Fig:borders}}
\end{figure}

\paragraph*{Covering probability.}
To obtain the covering probability for the points \emph{on} a boundary, we recall Eq.\ \eqref{eq:pinside} for the probability of having at least one covering point in a region of volume $V_R$, and obtain $p_{\mathrm{bound},d-1} = 1-\exp(\Theta/2)=1-(1-p_C)^{1/2}$ for the center of a half sphere on a $(d-1)$-dimensional boundary, $p_{\mathrm{bound},d-2} = 1-\exp(\Theta/4)=1-(1-p_C)^{1/4}$ for the center of a quarter sphere at a $(d-2)$-dimensional boundary, and so on.
It is only slightly harder to compute the covering probability for points in the bulk at distance $\rho$ from a $(d-1)$-dimensional hypersurface,
\begin{equation}
p_{\mathrm{bulk}}(\rho) = 1-\exp(-\Theta V^{\|}_d (-r,\rho,r)/V_d r^d) =1-(1-p_C)^{V^{\|}_d(-r,\rho,r)/V_d r^d}.
\end{equation}
This function is shown in the left panel of Fig.~\ref{Fig:borders} for $p_C = 95\%$ and $d = 2\mbox{--}20$.

There are two simple strategies to correct for boundary effects: extending the bulk covering across the boundary, and adding a supplementary covering on the boundary. We now examine each in turn.

\subsection{First strategy: Extend the bulk covering across the boundary}

Looking at the left panel of Fig.\ \ref{Fig:borders}, we see that the covering probability for points in the bulk decreases significantly only at distance $\rho < r/2$ from the $(d-1)$-dimensional boundary. If we choose to extend the bulk covering to a distance $r/2$ beyond the boundary, we see that the total number of covering points must increase by a factor $\simeq (1+(S/V) \times r/2)$, where $S$ is the hypersurface of the covered hypervolume $V$, at least for sufficiently smooth hypersurfaces. At any given dimension, the ratio $S/V$ achieves the minimum possible value for a hypersphere, where $S/V = d/R$ (for a hypercube, $S/V = 2d/L$). Thus the increase in the number of points incurred by extending the covering has a lower bound $\simeq (1 + d \, r/D)$, where $D$ is the linear spatial extent of the covered volume. The less symmetric the covered volume, the more important the boundary effects, which can be dominated by the dimension of shortest extent; of course if this extent is much smaller than the covering radius, the covering problem is effectively one of a lower dimension.

We tested these theoretical predictions \textit{\`a la} Monte Carlo, by repeatedly generating random coverings for the $d$-dimensional hypercube. For all the tests in this paper we employed a Mersenne-twister pseudorandom number generator \cite{mersenne}, which has extremely long period and is expected to generate uncorrelated sequences of points (i.e., $d$-uples of reals) up to 623 dimensions. (Linear congruential generators \cite{recipes}, by contrast, tend to generate $d$-uples that lie on hyperplanes, which would affect covering performance much as it does for periodic lattices.) We generated coverings of 
\begin{equation}
N_E = \Theta^\mathrm{random}_{95\%} (1+2\delta)^d\left( \frac{1}{V_d r^d} \right)
\end{equation}
random points, uniformly distributed in the hypercube $[-\delta,1+\delta]^d$ (with $\delta = r$ or $r/2$); we then estimated the covering fraction by drawing 10,000 (target-signal) points uniformly across the hypercube $[0,1]^d$. The resulting $N_E$ and covering fractions are shown in Table~\ref{test_copertura}. For each $d$, we chose $r$ so that a bulk-only covering would have between 100 and 100,000 points (which are representative values for the template banks used in GW searches). Clearly the number of points in the extension across the boundary grows dramatically with $d$.

This of course would be true also for periodic-lattice coverings, which have ``overflow.'' Consider for instance the covering of $[0,1]^8$ with $\mathbb{Z}^8$ and $r = 0.3$. Because
\begin{equation}
\Theta^{\mathbb{Z}^8}_{95\%} \left( \frac{1}{V_8 \times 0.3^8} \right) \simeq 4.53 \left( \frac{1}{4.059 \times 0.3^8} \right) \simeq 17,011 = (3.38)^8,
\end{equation}
we need to choose between a covering of $3^8 = 6,561$ points with poor coverage near the sides, or a much more expensive covering of $4^8=65,536$ points. Boundary effects are certainly less important when $r$ is smaller, but the overall numbers are much larger. For instance, for $r = 0.05$ we need to choose between 20 and 21 points to each side. Now, $(21/20)^8 \simeq 1.48$, but $20^8=2.56 \times 10^{10}$.
\begin{table}
\begin{tabular}{l|rr|rr|rr|rrc}
\hline \hline
$d$ &  \multicolumn{2}{c|}{Covering sphere} & \multicolumn{2}{c|}{Bulk + border, $\delta = r$} & \multicolumn{2}{c|}{Bulk + border, $\delta = r/2$} & \multicolumn{3}{c}{Bulk only} \\
    &   \multicolumn{1}{c}{$r$}      & \multicolumn{1}{c|}{$V_d r^d$} & \multicolumn{1}{c}{$N$} & \multicolumn{1}{c|}{cov.\%} & \multicolumn{1}{c}{$N$} & \multicolumn{1}{c|}{cov.\%} & \multicolumn{1}{c}{$N$} & \multicolumn{1}{c}{cov.\%} & cov.\% (theory) \\
\hline 
1   &    $10^{-3}$   & $2.00 \times 10^{-3}$     &  1,500   & 95\%     &   1,499  & 95\%      &    1,497     & 95\%    & 95\% \\
    &    0.01    & $2.00 \times 10^{-2}$     &    152   & 92--96\%  &     151  & 92--96\%   &      149      & 92--96\% &95\% \\	    			         				   
2   &    0.01    & $3.14 \times 10^{-4}$     &  9,920   & 95\%     &   9,727  & 95\%      &    9,535     & 95\% &95\%  \\
    &    0.05    & $7.85 \times 10^{-3}$     &    461 & $\simeq$ 95\% &  420 & $\simeq$ 95\%  &  381      & $\simeq$ 94\% &94\%  \\	    			         				   
3   &    0.05    & $5.24 \times 10^{-4}$     &  7,615   & 95\%     &   6,623  & 95\%      &    5,721    & 93\% &94\%  \\   
    &    0.10    & $4.19 \times 10^{-3}$     &  1,235   & 95\%     &     951  & 95\%      &      715     & 92\% &92\%  \\ 	    			         				   
4   &    0.05    & $3.08 \times 10^{-5}$     &142,207   & 95\%     & 118,062  & 95\%      &   97,129   & 93\% &93\%  \\           
    &    0.10    & $4.93 \times 10^{-4}$     & 12,588   & 95\%     &   8,887  & 95\%      &    6,070    & 91\% &92\%  \\           
    &    0.20    & $7.90 \times 10^{-3}$     &  1,457   & 95\%     &     786  & 95\%      &      379     & $\simeq$ 88\% &90\%  \\           	    			         				   
6   &    0.15    & $5.89 \times 10^{-5}$     &245,650   & 95\%     & 117,718  & 95\%      &   50,892   & 87\% &90\%  \\
    &    0.20    & $3.31 \times 10^{-4}$     & 68,201   & 95\%     &  27,046  & 95\%      &    9,057    & 84\% &89\%  \\
    &    0.30    & $3.77 \times 10^{-3}$     & 13,341   & 95\%     &   3,838  & 94\%      &      795     & 78\% &87\%  \\  	    			         				   
8   &    0.30    & $2.66 \times 10^{-4}$     &483,175   & 95\%     &  91,768  & 95\%      &   11,249   & 74\% &86\%  \\  
    &    0.40    & $2.66 \times 10^{-3}$     &124,112   & 95\%     &  16,621  & 95\%      &    1,126    & 66\% &85\%  \\  
    &    0.50    & $1.59 \times 10^{-2}$     & 48,372   & 95\%     &   4,842  & 93\%      &      188     & $\simeq$ 57\%  &84\%  \\  
\hline \hline
\end{tabular}
\caption{Number of additional templates and resulting covering fraction with the covering-extension strategy for a $[0,1]^d$ hypercube, as a function of dimension $d$ and covering-sphere radius $r$. The required covering fraction is 95\% for all runs. Columns 4 and 5 (6 and 7) show the number of covering points and the achieved covering fraction, to 1\% accuracy, for a border of width $\delta = r$ ($\delta = r/2$). Columns 8 and 9 show the same information for a bulk-only covering; column 10 shows the theoretical covering fraction for a bulk-only covering obtained by considering only $(d-1)$-dimensional effects, which suggests that $(d-2)$- and lower-dimensional boundaries become important as $d$ increases. See the Appendix for details about numerical methods.
\label{test_copertura}}
\end{table}

To apply the extension strategy on a curved manifold, we need to determine how far the covering must be extended in coordinate space to achieve the proper bordering, which requires knowledge of the metric. For GW template banks, we need also to worry that a waveform with parameters outside the original region of interest may not be physical, or may not even exist. For this reason algorithms such as the square-lattice placement of Ref.\ \cite{Babak:2006p1262} place templates on the boundary first, and then in the bulk. This is just the second strategy that we consider next.

\subsection{Second strategy: Add a lower-dimensional covering on the boundary}

This scheme is generally easier to implement than the extension across boundaries, and it is a natural fit for the metricless, mesh-based placement methods described later in this paper.
Now, how many points must be placed on a $(d-1)$-dimensional boundary to achieve uniform covering throughout the bulk? To answer this question, assume we lay down a boundary covering with covering probability $p_S$, and compute the probability of covering a generic bulk point with at least one sphere from the boundary covering:
\begin{equation}
p_{\mathrm{surf}}(\rho) = 1 - \left( 1 - p_S \right)^{(1-(\rho/r)^2)^{(d-1)/2}} \quad
\text{for $\rho < r$};
\end{equation}
this function is shown in the center panel of Fig.\ \ref{Fig:borders} for $p_S = 0.78$, and can be obtained easily by realizing 
that a $(d-1)$-dimensional boundary covering with radius $r$ becomes a covering with radius $r'$ on a $(d-1)$-dimensional surface parallel (at least locally) to the boundary. The new radius $r'$ can be determined using the Pythagorean theorem (see Fig.~\ref{Fig:correlations_borders}), and the relation between the covering probabilities using Eq.~\eqref{eq:pinside} with $V_R=V_{(d-1)} (r')^{(d-1)}$.

Close to the boundary, the combined covering probability from the bulk and the boundary coverings will then be
\begin{equation}
p_{\mathrm{bulk+surf}}(\rho)=1-\bigl(1-p_\mathrm{bulk}(\rho)\bigr)\bigl(1-p_{\mathrm{surf}}(\rho)\bigr).
\end{equation}
If we choose $p_S = 1-\exp(-\Theta/2)$, the points on the boundary get the same covering probability as those in the bulk, $p_{\mathrm{bulk+surf}}(0)=1-\exp(-\Theta)=p_C$.
Close to the boundary, 
\begin{equation}
p_{\mathrm{bulk+surf}}(\rho)=1-\exp(-\Theta\, C(\rho)), \quad \text{with} \quad
C(\rho)=V^{\|}_d(-r,\rho,r)/V_d r^d + \frac{1}{2}(1-(\rho/r)^2)^{(d-1)/2}.
\end{equation}
As expected, $p_\mathrm{bulk+surf}$ is $p_C$ for $\rho = r$, and larger for $0 < \rho < r$. 

In principle it is possible to apply the same strategy to boundaries of dimension lower than $d-1$, but the best arrangement will vary from case to case. Consider for example the 1-dimensional edges of a 3-dimensional cube: after placing a random covering in the bulk and on the 2-dimensional sides as we have just outlined, the covering probability on the edges is $1-(1-p_{3})(1-p_{2})^2=1-\exp(- (3/4) \Theta)$, where $p_{3}=1-\exp(-\Theta/4)$ is the probability of an edge point  being covered by at least one point from the bulk, while $p_{2}=1-\exp(-(\Theta/2)/2)$ is the probability of it being covered by a point from one of the sides shared by the edge. So an additional one-dimensional covering must be placed on the edge with $p_E = 1-\exp(-(\Theta/4))$ to achieve a covering probability $1-\exp(-\Theta)$ on the edges. Notice that to set $p_E$ we have used the information that two sides that share an edge are orthogonal, which will not be the case in general.
To apply the boundary-covering strategy on a curved manifold, we need to compute the determinant of the metric \emph{on the hypersurface} in terms of the $d-1$ parameters that define it.

How do the two strategies compare? For a $[0,l]^d$ hypercube, the points on the $(d-1)$-dimensional boundary covering would be
\begin{equation}
N_S = \left( \frac{\Theta}{2} \right) \frac{2 \, d \, l^{d-1}}{V_{d-1} r^{d-1}},
\end{equation}
to be compared with the extension points
\begin{equation}
N_E - N_\mathrm{bulk} = \Theta \frac{l^d(1+2(r/2)/l)^d -l^d}{V_d r^d} \simeq \Theta \frac{ d \, r \,  l^{d-1}}{V_d r^d}
\end{equation}
(where we assume $r \ll l$). Thus $N_E / N_S = V_{d-1} / V_{d}$, which is $\sim 1$ to very high $d$. This result is correct also for more general geometries, as long as the bordering volume scales as the hypersurface times $r/2$.
However, the two strategies will in general have different distributions of $r_w$ (the radius of the largest empty sphere that can be fit among the covering points), since these depend on $d$.

\section{Self-avoiding coverings}
\label{sec:alternatives}

As pointed out by Messenger and colleagues \cite{Messenger2009} and as discussed above,
for sufficiently high $d$, random partial coverings become more efficient (i.e., have lower $\Theta$) than the best-known periodic lattices. In addition, they conform naturally to curved manifold geometries, given only a knowledge of the metric's determinant. However, each point in a random covering is placed independently of every other point, so it is natural to ask whether random coverings can be improved by making them \emph{self-avoiding}, while preserving their random, unstructured character. In this section we consider two ways to introduce self-avoidance: the use of \emph{quasirandom sequences} and the \emph{stochastic} schemes that accept each random draw only after considering its distance from the points already accepted into the covering.

\subsection{Quasirandom sequences}

\begin{figure}[t]
\includegraphics[width=\textwidth]{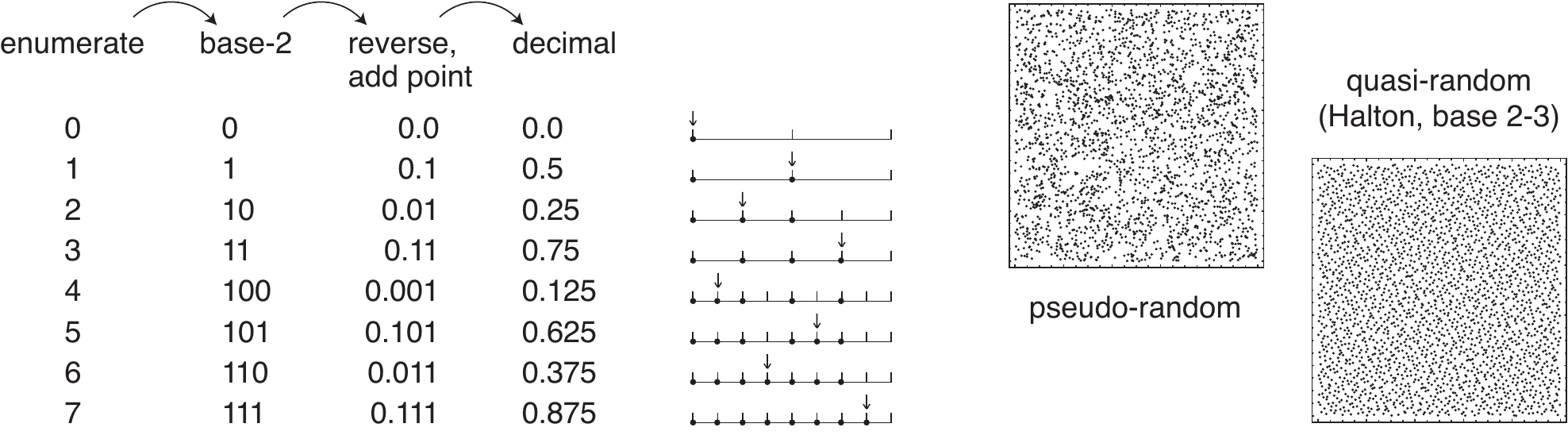}
\caption{Halton's quasirandom sequence. Left: to obtain the $n$-th number in the sequence, write $n$ as a base-$b$ number (here 2), reverse the digits, add a radix point at the left, and interpret the resulting string as a number $\in [0,1)$. This algorithm fills a succession of finer and finer Cartesian grids, spreading out the points maximally on each, since the most rapidly changing $b$-ary digit of $n$ controls the most significant digit of the placement \cite{recipes}. Right: the combination of base-2 and -3 Halton sequences fills the unit square more uniformly than the same number of (pseudo-)randomly placed points.
\label{fig:halton}}
\end{figure}

Quasirandom sequences, also known as low-discrepancy sequences \cite{recipes}, are designed to cover multidimensional regions more uniformly than pseudorandom $n$-tuples, although they may not appear as ``random'' (i.e., their deterministic nature is manifest). They are sometimes used to improve the convergence of multidimensional Monte Carlo integration. A conceptually simple example is Halton's sequence \cite{recipes}, shown in Fig.\ \ref{fig:halton}. Quasirandom sequences can be used as slot-in replacements for pseudorandom numbers in random coverings.

\begin{figure}[t]
\includegraphics{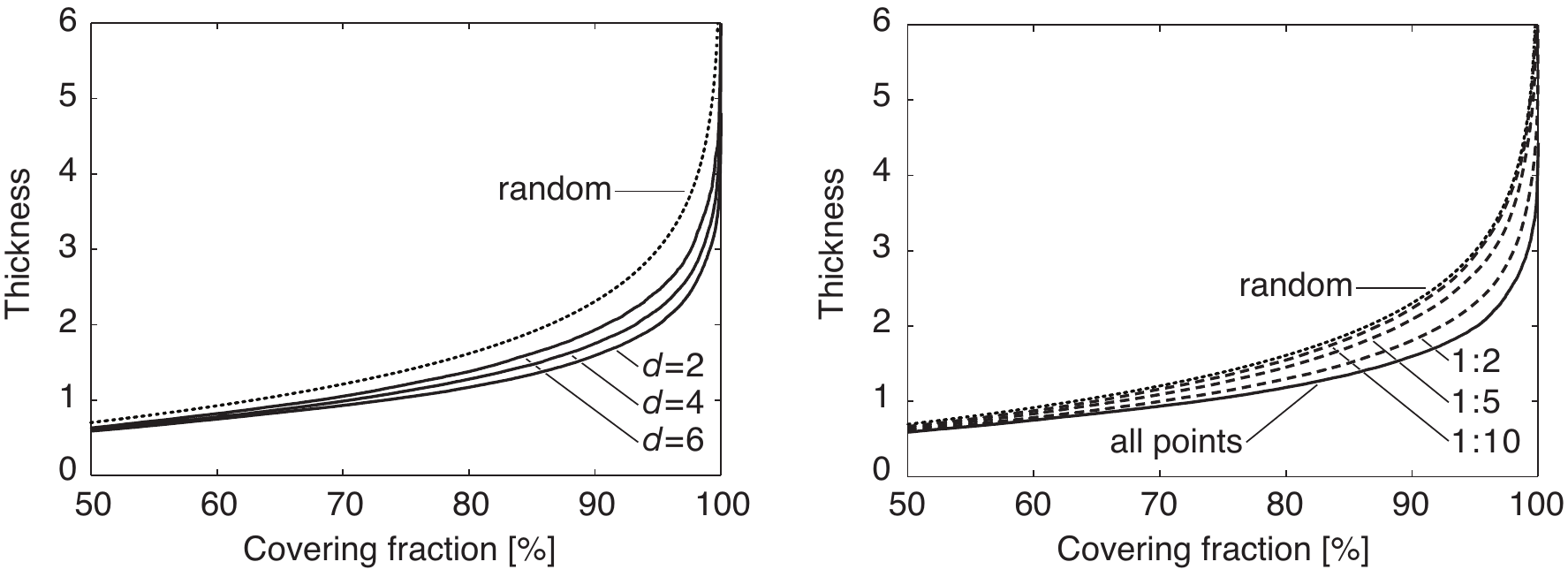}
\caption{Left: effective thickness as a function of covering fraction for Sobol quasirandom coverings of bulk $\mathbb{E}^d$, for $d =$ 2, 4 and 6, as compared to the theoretical expectation for random coverings. (As a control, we used the same ``experimental'' setup to evaluate the random-covering thickness for $d = $ 2, 4 and 6, which matches the theoretical expectation very accurately, and is not shown here.)
Right: deterioration of quasirandom-covering thickness on curved manifolds. We estimate the effect of applying a rejection method in the presence of a varying $\sqrt{|g_{kl}|}$ by extracting random subsets of 1/2, 1/5, and 1/10 of all points from a Sobol covering of $\mathbb{E}^2$, and evaluating the resulting thickness. Already for a range of variation $\sim 10$ in $\sqrt{|g_{kl}|}$, the thickness grows very close to its random-covering value for the same covering fraction.
\label{Fig.Sobol_reject}
\label{Fig.Sobol_dim_dil}}
\end{figure}

\paragraph*{Effective thickness.}
Numerical experimentation with the more sophisticated and widely adopted Sobol sequences \cite{sobol}, shows that their effective thickness in (bulk) $\mathbb{E}^d$ does significantly improve on the random-covering value, but it grows closer to the latter as $d$ increases (see Table \ref{Table:Sobol} and the left panel of Fig.\ \ref{Fig.Sobol_dim_dil}).
Indeed, the quasirandom thickness may approach the random-covering value asymptotically as $d \rightarrow \infty$, but a proof (or disproof) remains to be found. In addition, the thickness of quasirandom coverings appears relatively stable under dilations along one axis.
\begin{table}[t] 
\begin{tabular}{l|lll}
\hline \hline
$d$ & $\Theta_{90\%}$ & $\Theta_{95\%}$ & $\Theta_{99\%}$    \\
\hline
2 & 1.6 & 2.0 & 2.9  \\
3 & 1.7 & 2.1 & 3.0  \\
4 & 1.8 & 2.2 & 3.2  \\
6 & 1.9 & 2.5 & 3.6  \\
random & 2.3 & 3.0 & 4.6 \\
\hline \hline
\end{tabular}  
\caption{Numerically determined effective thickness of Sobol quasirandom coverings of bulk Euclidean space (no boundary effects). See Appendix for details on the numerical experiment.
The same setup was used for Fig.~\ref{Fig.Sobol_dim_dil}.}
\label{Table:Sobol}
\end{table}

\paragraph*{Curved manifolds and boundary effects.}
To use quasirandom coverings on curved manifolds, their density can be locally modulated in proportion to $\sqrt{|g_{kl}(\lambda_i)|}$, using a variant of the rejection method discussed in Sec.\ \ref{sec:random}. Unfortunately, such a process tends to destroy the self-avoidance of quasirandom sequences, resulting in effective thicknesses close to the random-covering values (see right panel of Fig.\ \ref{Fig.Sobol_reject}). Thus the utility of quasirandom coverings appears limited to flat or almost flat manifolds. As for boundary effects, the geometrical analysis of Sec.\ \ref{sec:boundary} still applies, so the same covering strategies can be carried over, using empirically determined relations between thickness and covering percentage such as those plotted in the left panel of Fig.~\ref{Fig.Sobol_dim_dil}.

\subsection{Stochastic placement}

\begin{table}
\begin{tabular}{ll|ll|ll|ll|ll}
\hline \hline
$d$ & $\gamma$ &        $\Theta_{80\%}$ & $\alpha_{80\%}$ & $\Theta_{90\%}$ & $\alpha_{90\%}$ & $\Theta_{95\%}$ & $\alpha_{95\%}$ & $\Theta_{99\%}$ & $\alpha_{99\%}$ \\
\hline
2 & 0.5      & 1.41 & 1.2 & 1.90 & 1.3 & 2.33 & 1.4 & 3.10 & 1.7 \\
  & 0.6      & 1.33 & 1.3 & 1.76 & 1.5 & 2.13 & 1.6 & 2.78 & 2.1 \\ 
  & 0.7      & 1.24 & 1.4 & 1.61 & 1.7 & 1.93 & 1.9 & 2.46 & 2.6 \\
  & 0.8      & 1.20 & 1.6 & 1.52 & 2.0 & 1.77 & 2.4 & 2.17 & 3.6 \\
  & 0.9      & 1.12 & 1.9 & 1.42 & 2.4 & 1.62 & 3.1 & 1.97 & 5.8 \\
  & 1.0      & 1.07 & 2.2 & 1.32 & 3.1 & 1.50 & 4.5 & 1.80 & $\approx 11$ \\
  & 1.1      & 1.02 & 2.7 & 1.24 & 4.4 & 1.41 & $\approx 8$ & 1.66 & $\approx 60$ \\
  \hline
3 & 0.8   & 1.3 & 1.5 & 1.7  & 1.8 & 2.1  & 2.2 & 2.8 & 3.0  \\  
  & 1.0   & 1.2 & 2.2 & 1.5  & 3.2 & 1.8  & 4.6 & 2.2 & $\approx 12$ \\
\hline
4 & 0.8   & 1.4 & 1.35 & 1.9  & 1.5 & 2.4  & 1.75 & 3.4 & 2.3 \\
  & 1.0   & 1.3 & 2.2 & 1.7   & 3.1 & 2.0  & 4.5 & 2.7 & $\approx 12$ \\
 \hline
random & 0 & 1.61 & 1 & 2.3 & 1 & 3.0 & 1 & 4.6 & 1 \\
\hline \hline
\end{tabular}
\caption{Effective thickness of stochastic coverings of bulk $\mathbb{E}^d$, for $d = 2, 3, 4$ and for different choices of covering fraction and acceptance radius $\gamma$. The factor $\alpha$ is the ratio between the number of proposed and accepted points. For comparison, the last row shows the effective thickness of random coverings, for which $\gamma$ and $\alpha$ are formally 0 and 1.
\label{test_copertura_stochastic}}
\end{table}

Another approach to enforcing self-avoidance in random coverings is to assemble a covering by accepting or rejecting each random draw depending on already accepted points; such algorithms have been called \emph{stochastic}. Harry and colleagues~\cite{Harry} describe a \emph{destructive} process whereby an overdense covering is first generated; a point $P$ is then chosen randomly from it, and all points closer to $P$ than a distance $\gamma \, r$ are removed (with $r$ the covering radius, and $\gamma \in [0,2]$); this is repeated until all remaining points have relative distances larger than $\gamma \, r$. The resulting covering has slightly better thickness than a random covering with the same covering fraction. By contrast, Babak \cite{babak2008} implements a \emph{constructive} process that begins drawing random templates, and rejects all new points that are closer than a given distance from already accepted points. In our analysis below we set this distance to the multiple $\gamma \, r$ of the covering radius. (An important difference between the methods of Refs.\ \cite{Harry} and \cite{babak2008} is that Harry and colleagues use distances computed from a local metric, whereas Babak uses $\Delta$ distances, so his covering spheres can span multiple disconnected regions across parameter space.) 

\paragraph*{Equivalence of destructive and constructive processes.}
The constructive and destructive processes are exactly equivalent: it is possible to formulate them in such a way that for the same input (a random sequence $\{P^{(j)}\}$ of $N_d$ candidate points) they would produce the same covering. Let us see why.
In the destructive process, let $\{P^{(j)}\}$ be the points in the overdense covering, and consider them for inclusion in the thinned covering in the order $j = 1, 2, \ldots, N_d$. In the constructive process, use the same order to evaluate each point for inclusion. In both cases, once a $P^{(j)}$ is accepted, its presence rules out accepting all $P^{(k)}$ (with $k > j$) closer than $\gamma \, r$ to $P^{(j)}$. It is immaterial whether all such $P^{(k)}$ are discarded immediately (as in the destructive process) or as they are ``called up'' (as in the constructive process). Thus the final set of accepted points is the same in both cases.
Furthermore, the set of all the distances that must be computed is the same: it consists of the distances between each accepted $P^{(k)}$ and all accepted $P^{(j)}$ with $k > j$, and of the distances between each discarded $P^{(k)}$ and all accepted $P^{(j)}$ up to the first $P^{(j')}$ closer than $\gamma \, r$.

\paragraph*{Effective thickness.}
The choice of $\gamma$ is important. Setting $\gamma = 2$ yields a solution of the sphere-\emph{packing} problem \cite{Conway1999}, so for $\gamma$ close to 2 not all covering fractions can be achieved, because there is a limit to the number of solid impenetrable spheres that can be fit in a given volume. A key quantity that can be determined empirically is the average number of draws $\alpha$ that are needed to accept a point: it is defined as $\alpha = N_d / N$, where $N_d$ is the total number of draws, and $N \simeq \Theta^\mathrm{stoch}_{X\%} V_m / (V_d r^d)$ is the number of points of the final covering. Table \ref{test_copertura_stochastic} shows the empirically estimated thickness and the $\alpha$ factor for stochastic coverings of bulk $\mathbb{E}^{2\mbox{--}4}$ at different values of $\gamma$. The thickness improves (but $\alpha$ increases) for higher $\gamma$; the gains are diminishing at higher covering fractions, and at higher $d$. This confirms the general tendency that in higher dimensions it becomes progressively harder to improve the thickness of pure random distributions.

Instead of keeping $\gamma$ constant, it is also possible to vary it while points are being added to the covering; in particular, by decreasing $\gamma \propto N_a^{-1/d}$ (where $N_a$ is the number of points accepted so far) we can keep the number of attempts needed to accept a new point roughly constant. 
This can be understood as follows. The probability of a new point of being accepted is proportional to the fraction of the manifold left uncovered by the covering spheres of radius $\gamma\, r$, and therefore to $1 - p_C$, where $p_C$ is the (average) covering fraction attained so far for a covering radius $\gamma \, r$. Since $N_a (\gamma \, r)^d V_d= \Theta(p_C) V_m$, the covering fraction can be kept constant by changing $\gamma$ so that $N_a \gamma^d$ remains constant. The initial value of $\gamma$ is arbitrary, but it must correspond to an initial sphere volume $(\gamma \, r)^d V_d \sim V_m$ if we are to produce a sequence with thickness $\Theta_{X\%}^{\mathrm{var} \gamma}$ appreciably different from a pure random covering.

The thickness of such coverings is slightly worse than the thickness of the constant-$\gamma$ coverings that yield the same average $\alpha$ (see Table \ref{tab:vargamma}). However, variable-$\gamma$ coverings are interesting because they are \emph{scale free}, just as random and quasirandom coverings: that is, they are independent of the covering radius used to build them. Thus they can be produced in advance and stored, and then used as a poor man's version of quasirandom sequences. To generate a covering with covering fraction $X\%$ and covering radius $r$, we would simply use the first $N = \Theta_{X\%}^{\mathrm{var} \gamma} V_m / (V_d r^d)$ points in the sequence. However, their thickness is not stable under dilations, but converges to the thickness of pure random coverings. 
\begin{table}[t]
\begin{tabular}{lr|l|l|l|l}
\hline
\hline
$d$ & $\alpha$ &    $\Theta_{80\%}$ & $\Theta_{90\%}$ & $\Theta_{95\%}$ & $\Theta_{99\%}$ \\
\hline
$2 \;$ & 2.0     & 1.15  & 1.57 & 1.99 & 2.91 \\
  & 3.0     & 1.05  & 1.42 & 1.74 & 2.50 \\
  & 4.0     & 1.01  & 1.35 & 1.66 & 2.32 \\
  & 10.0      & 0.94  & 1.19 & 1.43 & 1.89 \\
\hline
3 & 2.0     & 1.23 & 1.70 & 2.14 & 3.20 \\
  & 3.0     & 1.14 & 1.55 & 1.95 & 2.9 \\
  & 4.2     & 1.11 & 1.50 & 1.87 & 2.6 \\
  & 12.0      & 1.05 & 1.39 & 1.68 & 2.3 \\
\hline
4 & 2.0     & 1.3  & 1.8 & 2.3 & 3.4 \\
  & 4.5     & 1.2  & 1.7 & 2.1 & 3.0 \\
  & 15.0      & 1.2 &  1.6 & 2.0 & 2.8 \\
\hline \hline
\end{tabular}
\caption{Effective thickness of scale-free (variable-$\gamma$) stochastic coverings of bulk $\mathbb{E}^d$, for different choices of covering fraction and acceptance factor $\alpha$.\label{tab:vargamma}}
\end{table}

\paragraph*{Computational cost.}
In principle we obtain the greatest covering efficiency by setting $\gamma$ as high as allowed by the desired covering fraction, and accepting a correspondingly large $\alpha$.
In the practice of matched filtering, however, we must balance the computational cost of placing a template with the cost of using it to filter the detector data. The latter is proportional to $N$, but the former is proportional to the number of distances that need to be computed in the stochastic process, which is $\lesssim \alpha N^2/2$. (This estimate includes $N \times N/2$ distances between accepted points, and $(\alpha - 1) N \times N / 2$ distances between discarded and accepted points. The second number is an upper limit because each discarded point could have been eliminated after comparing it with one of several accepted points; but the number of these neighbors is generally small compared to $N$.)
The cost of computing distances may be negligible if these can be derived reliably from an analytic metric, or from a numerically obtained metric that is constant across $\mathcal{H}$. On the other hand, the cost may be considerable if each distance requires the actual generation of templates, either to take their $\Delta$ distance directly, or to compute numerical metrics at different points in $\mathcal{H}$.
If the placement cost is dominated by generating templates as opposed to computing with them (e.g., taking their $\Delta$ distances), then its scaling can be attenuated to $\sim \alpha N$ by storing all templates as they are generated. Alternatively, the total number of distances that need to be computed can be made to scale as $\sim d \, \alpha \, N$ if it is possible to compare source parameters to decide which templates are likely to be neighbors. 
It is also true that placement cost is not an issue when detectors are stable enough that noise can be considered stationary [remember that noise levels affect distances through Eq.\ \eqref{eq:innerproduct}]; in that case, the bank can be placed once and reused across a long dataset, so that the total computational cost would be dominated by the filtering.

\paragraph*{Curved manifolds.}
It would appear \emph{prima facie} that the rejection process implicit in stochastic algorithm could replace the local density modulation of random draws needed by random coverings to cover curved $\mathcal{H}$ uniformly. This is correct, but there are two caveats. First, the stochastic process can become very inefficient on curved manifolds, since the average number of draws needed to accept a point is not $\alpha$, but an average of $\alpha \times (\max_{\lambda_i} \sqrt{|g_{kl}(\lambda_i)|})/\sqrt{|g_{kl}(\lambda_i)|}$ across $\mathcal{H}$. Second, the covering fraction of the final covering will not be uniform, but it will slightly favor the regions where $\sqrt{|g_{kl}(\lambda_i)|}$ is lowest. Thus, if the determinant of the metric is available across $\mathcal{H}$, we recommend combining the stochastic process with the nonuniform generation of random points.

\section{metricless, mesh-based random coverings of curved manifolds}
\label{sec:triang}

\begin{figure}[t]
\includegraphics{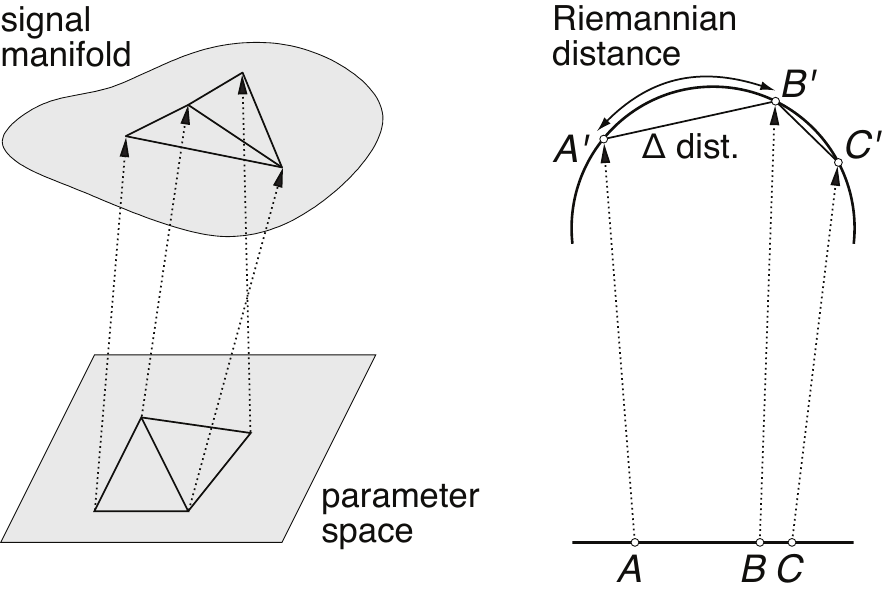}
\caption{Mapping of a parameter-space triangulation $T$ onto the signal manifold $\mathcal{H}$. If the $\Delta$ edge lengths are smaller than the characteristic curvature scale $\delta$ of the manifold, the image of $T$ will lie close to $\mathcal{H}$.}
\label{Fig:triangulation}
\end{figure}
Throughout this paper, we have assumed that we could transfer the covering strategies formulated for $\mathbb{E}^d$ to curved manifolds $\mathcal{H}$ using the pointwise knowledge of $\sqrt{|g_{jk}|}$. (A possible exception are ``straight'' stochastic coverings of curved curved manifolds, which however can be very inefficient, as discussed above.) We now turn to the case where $\sqrt{|g_{jk}|}$ is not available. In the context of template placement, this may happen when we do not have a (semi\mbox{-})analytical expression for the metric, because the waveform equations cannot be differentiated with respect to source parameters (e.g., the waveforms may be generated from the numerical solution of differential equations); the next recourse would be to take the numerical derivatives of Eq.\ \eqref{eq:distance} for small source-parameter displacements, but this too can prove difficult because of numerical noise or computational cost.

In this section we show how a discrete data structure consisting of a triangulation of coordinate (parameter) space and of the $\Delta$ distances measured along the triangulation's edges can be used \emph{in lieu} of a metric to build properly density-modulated random coverings. (By contrast, Beauville and colleagues \cite{Beauville:2003p1266} describe the use of a refined triangulation to interpolate equal-distance contours across two-dimensional parameter space to guide the placement of a locally hexagonal lattice.) 

\paragraph*{Triangulations.} To triangulate coordinate space, we decompose it into \emph{simplexes} (the $d$-dimensional analogues of triangles) such that their union covers the coordinate-space region of interest, and their intersection has zero $d$-dimensional volume. If the triangulation $T$ is dense enough that the Riemannian distances measured in $\mathcal{H}$ along the triangulation edges are smaller than a scale $\delta$ 
at which the curvature of $\mathcal{H}$ is negligible, then the image of $T$ in $\mathbb{R}^{N_\mathrm{big}}$ follows $\mathcal{H}$ closely (see Fig.\ \ref{Fig:triangulation}), and the Riemannian distances are approximated well by the $\Delta$ distances between neighboring vertices in $T$. In the following, we shall make this crucial assumption. We shall also assume that $\sqrt{|g_{jk}|}$ is almost constant across each triangle; this condition depends of course on the choice of coordinates as well as the intrinsic geometry of $\mathcal{H}$.

By construction, we may now approximate the volume of a region of $\mathcal{H}$ by the sum of the volumes of all simplexes within the region, as computed using the Cayley-Menger determinant formula~\cite{cayley-menger} and the $\Delta$ edge lengths. The triangular inequality \eqref{eq:triangle} for $\Delta$ guarantees that all simplexes have well-defined volumes. Furthermore, the ratio between the $\Delta$ volume of a simplex in $\mathcal{H}$ and its Euclidean volume in coordinate spaces approximates $\sqrt{|g_{jk}|}$, although in practice this calculation can suffer from numerical noise.

More formally, we see that the triangulation, together with the $\Delta$ edge lengths, carries the same information as the metric: if we assume that $g_{jk}$ is constant across a simplex, consistently with our assumption of local flatness, we can recover $g_{jk}$ by solving the distance equations
\begin{equation}
\Delta_A^2 = g_{jk} \Delta \ell_A^j \Delta \ell_A^k, \quad A = 1, \ldots, d(d+1)/2,
\label{eq:invertmetric}
\end{equation}
where $A$ enumerates the simplex edges, the $\Delta_A$ are their $\Delta$ lengths, and the $\Delta \ell^j_A$ are the components of vectors that lie along the edges in coordinate space.

\paragraph*{Random coverings.}
Thus armed with an estimate of the global volume of the region of interest in $\mathcal{H}$, and of the local $\sqrt{|g_{jk}|}$, we are now able to generate a properly density-modulated random covering of $\mathcal{H}$ by a rejection method. There is however a better algorithm that achieves the same result without discarding any random draw and without actually inverting Eq.\ \eqref{eq:invertmetric} to compute the full metric:
\begin{itemize}
\item To draw each new covering point, randomly select one of the simplexes in $T$ in such a way that the probability of choosing each simplex is proportional to its $\Delta$ volume, then pick a point randomly in coordinate space within the chosen simplex.
\item To select the simplex, form an $N_T$-dimensional vector given by the cumulative sum of the $\Delta$ volumes of all simplexes (arbitrarily ordered),
\begin{equation}
W_J = \sum_{I = 1}^{J} V_I,
\end{equation}
then draw a random number $x$ uniformly distributed in $[0,W_{N_T}]$, and choose the first simplex for which $W_J > x$.

\item To pick a point uniformly within the chosen simplex, simple algorithms such as the following can be used. Begin by considering the $d$-dimensional unit simplex, which in $\mathbb{R}^{d+1}$ is embedded in the $d$-dimensional hyperplane $y_1+y_2+\cdots+y_{d+1}=1$, and has vertex coordinates $y_a^b = \delta^b_a$ (for $a,b = 1,\ldots,d+1$). A random point in the unit simplex can be generated by drawing $d+1$ random numbers $r_a$ uniformly in $[0,1]$, and combining them as $x_a = \ln(r_a)/(\ln(r_1)+\cdots+\ln(r_{d+1}))$. This point can then be mapped to a point in an arbitrary $d$-dimensional simplex by the affine transformation $\lambda_j = x_1 v^1_j + \cdots + x_{d+1} v^{d+1}_j$, where the $v^a_j$ are the coordinates of the $a$-th vertex of that simplex. For details see Refs.\ \cite{simplex1,simplex2}.
\end{itemize}
Locally, this random covering is uniform: the assumption that the coordinates are distorted only by an affine transformation on the typical length-scale of a simplex edge ensures that $\sqrt{|g_{jk}|}$ is locally constant. Globally, by drawing points with density proportional to the $\Delta$ volume of simplexes, we correct for the variation of $\sqrt{|g_{jk}|}$ at larger scales. We can even estimate how many points we should draw to achieve a covering $X\%$ for covering radius $r$: from Eq.\ \eqref{eq:thickness} that is of course $N = \Theta_{X\%}^\mathrm{random} W_{N_T} / (V_d r^d)$.
To correct for boundary effects, we may generate boundary coverings guided by the lower-dimensional triangulation consisting of the faces of the full-dimensional simplexes that lie on the boundary.

\paragraph*{Generating triangulations.}
It is of course pointless to discuss the advantages of triangulation-guided coverings if we cannot prescribe a convenient procedure to build appropriately dense triangulations of coordinate space. 
We propose a solution based on the incremental refinement of an initial sparse triangulation, which may be generated randomly. The refinement can be stopped when the triangulation satisfies a criterion based on the $\Delta$ edge lengths (for instance, if we have an estimate of the curvature scale $\delta$, we may require that all edge lengths be safely below it). The scale $\delta$ can be seen as a tuning parameter: the validity of an estimate for $\delta$ can be checked \textit{a posteriori} by measuring the covering fraction of the final coverings and its variation across parameter space.

\begin{figure}[t]
\includegraphics{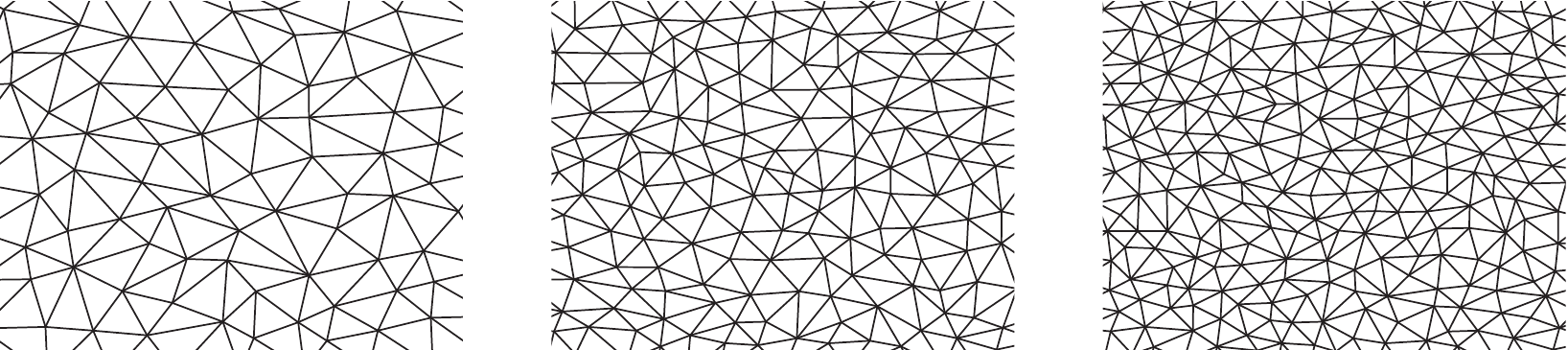}
\caption{
Three phases in the refined triangulation of $\mathbb{E}^2$, performed by placing new points at the barycenters of existing triangles.\label{Fig:bari}}
\end{figure}

The actual details of the refinement process depend on what \emph{kind} of triangulation we maintain. Consider for instance the \emph{Delaunay triangulation} of a set of points~\cite{recipes}. This is the unique triangulation such that the circumsphere of each simplex contains no other point; it has the property of (maximally) avoiding ``skinny'' simplexes with small angles, and therefore it is a good choice to model terrains (for $d=2$) or other hypersurfaces given a set of sample points. To refine a Delaunay triangulation, we can iteratively choose one of its simplexes on the basis of its $\Delta$ edge lengths, or of its $\Delta$ volume, place a new point at the barycenter, or randomly within the simplex, and retriangulate
(see Fig.~\ref{Fig:bari}).
Efficient \emph{incremental} algorithms exist that can adjust the triangulation to link the new point while preserving the Delaunay property (see, e.g., \cite{recipes}).

We have experimented with modifying these algorithms so that they would create triangulations that are Delaunay with respect to $\Delta$-wise (rather than coordinatewise) circumspheres. Such triangulations minimize the number of simplexes needed to approximate $\mathcal{H}$ faithfully 
but they suffer from a chicken-and-egg problem, because the local assessment of the Delaunay property is only reliable when the $\Delta$ edge lengths are already below the curvature scale $\delta$, which was the whole point of refining the triangulation. 
An alternative to iterative Delaunay triangulations are \emph{longest-edge} partition algorithms \cite{rivara}, 
which refine an initial triangulation by iteratively placing a new point on the longest one-dimensional edge, and dividing all the simplexes that share that edge. Again it may be useful to evaluate the longest edge with respect to $\Delta$ distances. 

\paragraph*{Refined-triangulation coverings.}
We note that it may be possible to use the very points of a refined triangulation as the points of a covering: in this case we would want to stop the refinement by comparing the population of edge lengths with the covering radius $r$. We investigate such algorithms in a separate paper \cite{vallismeshes}, but we note here that the numbers of edges in a triangulation (and thus the number of $\Delta$ distances to compute) grows very rapidly with $d$,
so triangulation-guided random coverings are still the better option if $\delta \gg r$ and if coordinates can be found such that the variation of $\sqrt{|g_{jk}|}$ across scales $\sim \delta$ is small.
If $\delta \sim r$, both triangulation-based approaches would still require fewer distance computations than ``straight'' stochastic placement, but the additional bookkeeping needed for the triangulation itself could become overwhelming, especially in higher dimensions.

\section{Conclusions and future prospects}
\label{sec:conclusions}

Past and current searches for modeled GW sources have largely relied on filtering detector data with carefully distributed banks of theoretical signal templates. Furthermore, even bankless Monte Carlo searches (as envisaged for the space-based detector LISA \cite{Vallisneri09}) can benefit from the exhaustive \textit{a priori} modeling of the posterior probability surface made possible by homogeneously distributed banks.
The notion of a Riemannian metric in parameter space \cite{Balasubramanian:1996p1260,Owen:1996p1258,Owen:1999p1259} allows placement methods based on periodic lattices \cite{Babak:2006p1262,Cokelaer:2007p1261}, which however are limited in practice to simple signal models with very few source parameters. Thus, future searches could greatly benefit from more generic and robust placement methods that are suited to signal models with complex parameter dependencies and with moderate number of parameters.

Template placement for generic signal families can be seen as an instance of the sphere-covering problem in Euclidean and Riemannian spaces. Working from this angle, Messenger and colleagues \cite{Messenger2009} examined the promise of \emph{random} coverings, while Babak \cite{babak2008} and Harry et al.\ \cite{Harry} studied \emph{stochastic} coverings that combine random draws with the enforcement of a minimum distance between pairs of points. In this article we have developed a deeper understanding of both kinds of coverings: specifically, we have derived analytically the variance of the covering fraction for random coverings, and examined the effects of boundaries; we have studied the \emph{self-avoiding} coverings generated by \emph{quasirandom} sequences; we have proved the equivalence of Harry et al.'s \emph{destructive} stochastic coverings with Babak's \emph{constructive} variant, and considered their effective thickness and computational cost; last, we have proposed a general technique to distribute coverings on curved signal manifolds using only the distances between the points of a parameter-space \emph{triangulation}, removing the need for the Riemannian metric, which may be difficult to obtain. 

Overall, our study confirms that randomized (random and stochastic) coverings compare very favorably to lattice-based coverings, especially for higher-dimensional parameter spaces, where randomized coverings provide greater simplicity and flexibility with comparable thicknesses. Furthermore, unlike lattices, randomized coverings generalize straightforwardly from Euclidean to Riemannian signal manifolds; the required modulation of local density can be achieved by computing the determinant of the metric, but also by metricless methods such as ``straight'' stochastic algorithms that compare $\Delta$ distances, and by triangulation-based algorithms.

Indeed, stochastic and triangulation approaches may be combined fruitfully: if the signal manifold has significant \emph{foldings} (i.e., distinct parameter-space regions that correspond to very similar signals with small $\Delta$ distances), a sufficiently refined triangulation-based covering would separately populate each duplicate region, and (as shown by Babak \cite{babak2008}) a subsequent stochastic stage could recognize the foldings and generate a list of nonlocal mappings. Such a list would be a very useful input to Monte Carlo searches that need to jump between isolated peaks on the likelihood surface \cite{Vallisneri09}. Furthermore, as pointed out by Babak \cite{babak2009}, the initial stage of Monte Carlo searches (before the chains latch onto a candidate signal) can be seen as filtering by yet another flavor of random banks, so some of the methods and estimates developed in this paper and in Refs.\ \cite{Messenger2009,Harry}, as well as our discussion of boundary effects, could be useful in that context.

Among the topics that we would like to flag for future investigation are the distance statistics of random and stochastic coverings, and especially the distribution of the maximum distance $\Delta_\mathrm{max}$; the possible mitigation of boundary effects with stochastic coverings, which naturally overpopulate the bulk regions near the boundaries; and the broad class of triangulation-based algorithms. It would also be interesting to investigate whether the interpolation of SNR across lattice-based template banks \cite{Croce2000,Croce2001,Mitra2005} can be extended to the products of randomized placement. 

As a final message, we wish to convey our belief, formed through the numerical experimentation carried out for this work, that the holy grail of a generally applicable template-placement algorithm is likely to remain unattainable: even general strategies such as random and stochastic coverings must be chosen, adapted, and carefully tuned for the specific search at hand. In every case, we first need to ask: which signals? what noise? what computational resources? The answer to these questions will guide the solution of what is arguably a problem of engineering applied to science.

\acknowledgments
We are grateful to B.\ Allen, S.\ Babak, D.\ A.\ Brown, T.\ Creighton, I.\ W.\ Harry, B.\ Krishnan, C.\ Messenger, B.\ J.\ Owen, I.\ Pinto, R.\ Prix, and B.\ S.\ Sathyaprakash for useful discussions. G.M.M.\ gratefully acknowledges support from the NASA postdoctoral program administered by Oak Ridge Associated Universities; M.V.\ from JPL's HRDF and RTD programs. This work was carried out at the Jet Propulsion Laboratory, California Institute of Technology, under contract with the National Aeronautics and Space Administration. Copyright 2009 California Institute of Technology. Government sponsorship acknowledged.

\appendix

\section{Numerical procedures}
\label{sec:numerics}

In this section we briefly describe the numerical procedures used to empirically determine covering fractions and effective thicknesses throughout this paper.

\paragraph*{Computing the thickness of partial periodic-lattice coverings.}
We work with the Voronoi cell of a lattice vertex, defined as the locus of points that are closest to that vertex than to any other (e.g., a hexagon centered on each vertex for the hexagonal lattice). We cover the Voronoi cell with a uniform distribution of points, and collect the vector $\Delta_i$ of their distances from the center. The $X$-th percentile of $\Delta_i$ is then the radius of the sphere covering that achieves a covering fraction of $X\%$, and the corresponding thickness is
\begin{equation}
\Theta_{X\%} = V_d (\Delta_{X\%})^d / V_\mathrm{cell},
\end{equation}
where the Voronoi-cell volume $V_\mathrm{cell}$ is given by the determinant of the generator matrix \cite{Conway1999}. 
This technique was used to generate the numbers of Table \ref{Tab:thickness} and Figs.\ \ref{Fig:partial_An} and \ref{Fig:dilatation_voronoi}.

\paragraph*{Verifying the covering fraction of random coverings with boundary effects.} As we do in the other two procedures described below, we begin by laying a very dense set of $M$ target points uniformly distributed across the region of $\mathbb{E}^d$ to be covered, typically the hypercube $[0,1]^d$ (which of course has $V_m = 1$). We then place a random covering of $N$ points throughout $[0,1]^d$, or throughout the larger hypercube $[-\delta,1+\delta]^d$, with $\delta = r/2$ or $r$, and verify what fraction of target points are covered (i.e., lie at a distance $< r$ from the closest covering point). This technique was used to generate Table \ref{test_copertura}.

\paragraph*{Computing the thickness of quasirandom coverings in the bulk.} Again we lay a dense target set in $[0,1]^d$; we then place a covering of $N$ points throughout a larger region that contains the hypercube. We compute the vector $\Delta_i$ of distances from the target points to the nearest point in the covering, and define
\begin{equation}
\Theta_i = N' V_d (\Delta_i)^d / V_m,
\label{eq:thick}
\end{equation} 
where $N'$ is the number of covering points that fall within $[0,1]^d$. By definition, Eq.\ \eqref{eq:thick} gives the thickness of a covering of $V_m$ with $N'$ points and covering radius $\Delta_i$; by our very experiment, if we set $\Delta_i$ to its $X$-th percentile, such a covering achieves a covering fraction equal to $X\%$. The $X$-th percentile of $\Theta_i$ is then the thickness of that covering. Boundary effects are avoided if the covering region is larger than $[0,1]^d$ by at least $\max_i \Delta_i$ on every side. This technique was used to generate the numbers of Table \ref{Table:Sobol} and Fig.\ \ref{Fig.Sobol_reject}.

\paragraph*{Computing the thickness of stochastic coverings in the bulk.}
We lay a dense target set on $[0,1]^d$, pick $r$ and $\gamma$, and place a stochastic covering with those parameters over a larger region that contains the hypercube.
We then examine the covering points one by one, and keep a running tally $C_k$ of the number of target points that have been covered by the first $k$ covering points. The thickness for covering fraction $X\%$ is obtained from
\begin{equation}
\Theta_{X\%} = k' V_d r^d / V_m,
\end{equation}
where we find the $k$ such that $C_k$ is $X\%$ of $M$, and set $k'$ equal to the number of covering points (among the first $k$) that fall within the hypercube. This technique was used to generate the numbers of Table \ref{test_copertura_stochastic}. (By contrast, the technique described in the paragraph above was used for the scale-free stochastic covering of Table \ref{tab:vargamma}.)

The last two techniques are equivalent, except that the first requires the \textit{a priori} choice of the number of covering points, the second of the covering radius.


\end{document}